\documentclass[12pt]{article}
\RequirePackage[OT1]{fontenc}
\RequirePackage{amsthm,amsmath}
\RequirePackage[colorlinks,citecolor=blue,urlcolor=blue]{hyperref}

 \usepackage[numbers]{natbib}
 \usepackage{amsmath,amssymb}
 \usepackage{amsthm}
 \usepackage{bm}
 \usepackage{latexsym}
 \usepackage{CJK}
 \usepackage{multirow}
 \usepackage{caption}
 \usepackage{graphicx,color}
 \usepackage{amssymb,amsfonts}
 \usepackage[english]{babel}
 \usepackage{times}
 \usepackage[T1]{fontenc}
 \usepackage{enumerate}
 \usepackage{algorithmic}
 \usepackage{algorithm}
\usepackage{geometry}
\usepackage{float}
\usepackage{appendix}
\usepackage{booktabs}
\usepackage{endnotes}
\usepackage{subcaption}
\usepackage{multicol}
\usepackage{tabularx}
\usepackage{lipsum}
\usepackage{setspace}
\newcommand{\blind}{1}

\newcommand{\argmin}{\arg\!\min}
\newtheorem{thm}{Theorem}
\newtheorem{lemma}{Lemma}
\newtheorem{prop}{Proposition}
\newtheorem{corollary}{Corollary}
\newtheorem{remark}{Remark}

\def\cL{\mathcal{L}}
\def\bA{\bm{\alpha}}
\def\bB{\bm{\beta}}
\def\bG{\bm{\gamma}}
\def\bN{\bm{\nu}}
\def\bL{\bm{\Lambda}}
\def\vec{\mbox{vec}}
\def\bd{\mbox{bdiag}}
\def\mG{\mathcal{G}}
\def\bT{\bm{\theta}}
\def\bD{\bm{D}_{N,m}}
\def\bH{\bm{H}_{N,m}}

\begin{document}

\def\spacingset#1{\renewcommand{\baselinestretch}%
{#1}\small\normalsize} \spacingset{1}

\if1\blind
{
  \title{\bf Individualized Multi-directional Variable Selection}
  \author{Xiwei Tang%\thanks{
    %The authors gratefully acknowledge \textit{please remember to list all relevant funding sources in the unblinded version}}\hspace{.2cm}
    \\
    Department of Statistics, University of Virginia\\
         \\
    Fei Xue \\
    Department of Statistics, University of Illinois Urbana and Champaign\\
         \\
    Annie Qu \\
    Department of Statistics, University of Illinois Urbana and Champaign}
      \date{}
  \maketitle
} \fi

\if0\blind
{
  \bigskip
  \bigskip
  \bigskip
  \begin{center}
    {\LARGE\bf Individualized Multi-directional Variable Selection}
\end{center}
  \medskip
} \fi

\bigskip

\begin{abstract}
In this paper we propose a heterogeneous modeling framework which achieves  individual-wise feature selection and  individualized covariates' effects  subgrouping simultaneously.
In contrast to conventional model selection approaches,  the new approach constructs a separation penalty  with multi-directional  shrinkages,
which facilitates  individualized modeling to distinguish strong signals from noisy ones  and  selects different relevant variables for different individuals.
Meanwhile,  the proposed model identifies subgroups among which individuals share similar covariates' effects,  and thus  improves  individualized estimation efficiency and feature selection accuracy.    Moreover,  the proposed model also incorporates    within-individual correlation for longitudinal data to gain extra efficiency.
We provide a general theoretical foundation under a  double-divergence modeling  framework where the number of individuals and  the number of individual-wise  measurements can both diverge, which enables  inference on both an individual level and a population level.
In particular, we establish  strong oracle property for the individualized estimator  to ensure its optimal large sample property under various conditions.  An efficient ADMM algorithm is developed  for computational scalability.   Simulation studies and  applications to  post-trauma mental disorder analysis with genetic variation and  an HIV longitudinal   treatment  study are illustrated to compare  the new approach to existing  methods.

\end{abstract}

\noindent%
\vfill

\newpage
\spacingset{1.5} % DON'T change the spacing!

\section{Introduction}
\label{sec:intro}

%{\qu
%% Individualized modeling (double-divergence)
 In recent years  there has been a growing demand  for exploring  individualized modeling, which has broad applications in personalized   medicine,  personalized education and personalized  marketing.      The traditional one-model-fits-the-whole-population  approach is unable  to detect important patterns and make  personalized  predictions for  specific individuals.
For example, in a genetic study to identify biomarkers associated with a certain disease, one gene could be a relevant biomarker for a subgroup of individuals in the population, but might not be a relevant biomarker for other individuals.
 Furthermore, the subgroup structure regarding heterogeneous covariates' effects might vary for different genes.
 Hence, individualized variable selection is very important as different individuals could have different sets of biomarker genes.  In addition, the rise of precision medicine and personalized marketing strategies also motivate  us to develop  more effective  personalized treatment and recommendation by selecting unique features for each individual.
 The  collection of  rich data information makes it feasible and compelling to utilize   individualized  models,  as   traditional   population models   cannot incorporate  heterogeneous  effects from different individuals. Therefore  it is  urgently needed   to  develop  new statistical methodology and theory  for variable selection and estimation for  individualized modeling.

%%Penalized variable selection
In the past two decades,    penalized model selection methods have been developed, e.g.,  the Lasso \citep{Tibshirani:1996}, the smoothly clipped absolute deviation (SCAD) \citep{Fan:2001}, the elastic net \citep{Zou:2005}, the adaptive Lasso \citep{Zou:2006},  the group Lasso \citep{Yuan:2006},  the minimax concave penalty (MCP)  \citep{Zhang:2010} and the truncated $L_1$-penalty (TLP) \citep{Shen:2012}.   %However,  the above methods are based on a homogeneous model setting which selects predictors for  entire populations.
One unique challenge of  individualized model selection  is that there could be  different relevant  or important predictors for different individuals.
A naive choice is to employ traditional  variable selection methods separately for  each individual,   if there are  multiple observations from each individual,  as in  longitudinal data settings.    However,  in practice,  the number  of  measurements for particular individuals could be limited.    In addition,  it is likely that some variables are  %population-shared  and are
invariant  for  the same individual,  such as demographic information variables, e.g.,    race and gender,    which  impose   restrictions and additional  obstacles to performing  individualized  variable  selection  based on  a standard  individual-wise  model framework.
%%Subgroupping of coefficients
Another  limitation  of applying   standard individual-wise  variable selection  is that it ignores information from other individuals which  might share similar effects on  important predictors of interest.     %Moreover,  assuming  each individual   to have    unique  effects for  all covariates   is practically unrealistic  and computationally  infeasible.
It  is more  sensible to  assume that  subpopulations of individuals  share common effects on  selected predictors.  Furthermore,  borrowing  information from  homogeneous subgroups allows one to  increase  estimation efficiency and  model selection accuracy.  % in addition to avoid over-fitting the model.   % Furthermore,  we allow  a subset of covariates  as population-shared predictors  if they are associated with the responses in a homogeneous  fashion for the  entire population.

In order to utilize cross-individual information,  we pursue  an underlying subpopulation structure depending  on   unobserved covariates.
Existing approaches dealing with clustering on regression coefficients include  mixture modeling for regression,  such as the mixture-of-experts model \citep{Jacobs:1991}.
 However,  most model selection approaches under this framework including  \citep{Raftery:2006}, \citep{Pan:2006} and   \citep{Guo:2010}   only  focus on  choosing  informative variables  to  distinguish  different subgroups,  rather than on selecting relevant predictors for different individuals.

Alternative approaches to model-based clustering on regression coefficients employ grouping penalization.  For example, \citep{Tibshirani:2005} propose a fused Lasso by adding an $L_1$-penalty  to the pair of adjacent coefficients;    \citep{Bondell:2008} propose  a  clustering algorithm for regression by imposing a  special octagonal shrinkage penalty on each pair of coefficients;  \citep{Shen:2010} develop a grouping pursuit algorithm utilizing the truncated $L_1$-penalty for fusions, and \citep{Ke:2013} propose  a  data-driven segmentation method to explore  homogeneous groups with regression.   Nevertheless, these are  all still  under the  population-regression model,  and  do not  allow  different individuals to have different features.     For the purpose of subgrouping different individuals, \citep{Hocking:2011} and \citep{Lindsten:2011} formulate  clustering    as a penalized regression problem by adopting an $L_p$-fusion penalty.    % \citep{Pan:2013}  show that the convex fusion-type  penalties  are   pairwise exhaustive,  which could lead to   biased estimations.
 \citep{Pan:2013} and \citep{Ma:2016} apply   non-convex fusion penalties to solve the bias problem.   However,  the fusion-type of penalty focuses on  subgrouping rather than  on model selection for individual coefficients.   % Indeed,   the aforementioned  subgrouping  approaches are not in the same direction with us and belong to a different topic. }

%% The proposed method

In this paper we propose an effective  individualized model selection approach utilizing multi-directional shrinkage to select unique  relevant variables for different individuals and identify subgroups based on heterogeneous covariates' effects simultaneously. To the  best of our knowledge,   this  is a new approach which has not been   offered  in the existing literature.
In the feature selection point of view, the proposed penalty allows  multiple possible  shrinking directions including the one  towards zero,   which differs  from conventional   penalty functions  with shrinking direction  towards zero only.
The consequence of conventional convex penalty functions (e.g., $L_p$-penalty) is that non-zero signals could suffer from zero-directional shrinkage, although a variety of penalty  methods have been proposed to solve the bias problem such as  non-concave penalties (e.g., SCAD,  MCP and TLP) or  adaptive weights (e.g., adaptive Lasso).  Instead  we propose a rather different approach  which shrinks different penalized parameters to different directions,  where the best shrinking option  is determined by the data itself.   One advantage of the proposed  method is that,  as long as the candidate directions contain  the one closest  to the truth,    the optimal large sample properties  such as the oracle property hold  by applying  a regular  $L_1$-type of  penalty in each direction.

In addition to individual-wise feature selection, our  paper considers a new  covariate-specific  subgrouping framework which is different from  traditional subgroup analysis in terms of the following: (1) pursuing subgroups based on heterogeneous covariates'  effects and allowing  subgrouping on individuals to vary over different covariates;  and (2) identifying the subgroup with null effects specifically,  which enables feature selection on an individual level.
Note that it is crucial to achieve  simultaneous feature selection and subgrouping,  as   post-subgrouping inference could suffer from potential estimation bias \citep{Desai:2014,Foster:2011}.
Moreover,  through utilizing  cross-individual information,   the proposed model  improves  estimation efficiency  and thus enhances  personalized prediction power.

%% Theory for DD
In theory,  we lay out  a theoretical framework for the double-divergence heterogeneous model with correlated data.     \citep{Xie:2003} and \citep{Balan:2005} established rigorous large sample theory for the  generalized estimating equation \citep{Liang:1986} (GEE)   estimator when the number of subjects  and the repeated measurement size are both large under a homogeneous setting;   and   \citep{Wang:2012}  investigate  the  GEE model with  high-dimensional  covariates,  but bounded repeated measurement size.  In this paper, we establish theoretical  properties  in a heterogeneous framework where the number of individuals and the individual-wise measurement size  are both increasing, which  involves  high-dimensional parameters as the number of individualized parameters is also increasing.  Furthermore, we develop   asymptotic  theory  for the proposed estimator under a variety of conditions and  establish  the optimal strong  oracle property for individualized model estimation and feature selection, and  uniform subgroup identification consistency.

The major contributions of  theory  development in this paper can be  outlined as follows. (1) Traditional subgroup analysis mostly  establishes theoretical results on the population or subpopulation level, for example, the average effect from a subgroup.   In contrast, the theoretical framework established  in this paper  provides an individual-wise model inference,  with a strong oracle property ensuring  optimal model selection consistency, estimation efficiency and subgroup identification consistency for each individual.
(2) To the best of our knowledge, in order to achieve the desired oracle property for either heterogeneous model estimation or  uniform subgroup identification consistency (all individuals correctly classified),  most existing penalization-based subgroup analyses  \citep{Tang:2017, Zhu:2018} consider the scenario of a fixed number of individuals $N$ and   a divergent number of measurements on each individual $m$, which could be restrictive in practice.  In contrast, the proposed double-divergence framework allows both $N$ and $m$ to diverge, which also provides the divergence rate of individualized parameters with respect to a divergent $N$.
(3) We also incorporate  within-individual correlation in the proposed model, and establish theoretical properties under mild conditions.  In fact, incorporating  individual-wise correlation brings non-trivial theoretical challenges  to the double-divergence framework since the dimension of the correlation structure diverges as individual measurement size $m$ increases.

%% Outline
The paper  is organized as follows. Section 2 introduces the general framework and presents the methodology.   Section 3 establishes  the theoretical results.  Section 4 discusses the computation and proposes an efficient algorithm.  Section 5 presents  simulation studies. Section 6 illustrates  an application on post-trauma mental disorder analysis from  the Detroit Neighborhood Health Study.  The last section provides concluding remarks and discussion.

\section{Model Framework and Methodology}

\subsection{Heterogeneous regression model}
We formulate  the problem under  the longitudinal data setting,  where each individual can have  multiple observations. 
For the $i$th individual,  let $\bm{y_i}=({y_{i,1}}, \ldots, {y_{i,m_i}})^T$ be an $m_i$-dimensional response variable, $\bm{X_i}=(\bm{x_{i1}}, \dots, \bm{x_{ip}})$ be an $m_i \times p$ covariate  matrix of predictors with heterogeneous effects,   and $\bm{Z_i}=(\bm{z_{i1}},\ldots,\bm{z_{iq}})$ be an $m_i \times q$ covariate matrix of  population-shared predictors.     We consider a heterogeneous  regression model:
\begin{equation*}
\setlength{\abovedisplayskip}{5pt}
\setlength{\belowdisplayskip}{7pt}
\bm{y_i}=\bm{X_i}\bm{\beta_i}+\bm{Z_i}\bm{\alpha}+\bm{\varepsilon_i}, \quad i=1,\ldots,N,
\end{equation*}
where each individual is associated with a unique effect $\bm{\beta_i}=(\beta_{i1},\ldots, \beta_{ip})^T_{p\times1}$ for some targeting variables $\bm{X}_i$,  in addition to  a  homogeneous effect $\bm{\alpha}=(\alpha_{1},\ldots, \alpha_{q})^T_{q\times1}$ for some control variables $\bm{Z}_i$.   The random errors $\bm{\varepsilon}_i=(\varepsilon_{i,1},\ldots,\varepsilon_{i,m})^T_{m\times1}$  are  independent over different individuals, while within an individual,  $\varepsilon_{i,t}$'s ($t=1,\ldots,m$)  have  mean 0 and variance $\sigma^2$,  and could be correlated.  For ease of notation, we assume a  balanced dataset  with $m_i=m$ in this paper.

In general, to identify unique features for different individuals, with an independent error assumption and a squared loss, we could employ a penalization method to select and estimate the regression parameters $\bm{\beta_i}$'s and $\bm{\alpha}$  through minimizing the penalized objective function 
\begin{equation}\label{objective:joint}
\setlength{\abovedisplayskip}{5pt}
\setlength{\belowdisplayskip}{7pt}
\frac{1}{2} \sum_{i=1}^N  \parallel \bm{y_i}- \bm{X_i}\bm{\beta_i}-\bm{Z_i}\bm{\alpha}\parallel_2^2+  \sum_{i=1}^N \sum_{k=1}^p h_{\lambda_{N,m}}(\beta_{ik}),
\end{equation}
%\begin{equation}\label{objective:general}
%\setlength{\abovedisplayskip}{5pt}
%\setlength{\belowdisplayskip}{7pt}
% (\hat{\bm{\beta}},\hat{\bm{\alpha}})=\argmin_{\bm{\beta},\bm{\alpha}}  \frac{1}{2}  \sum_{i=1}^N  L(\bm{y_i}- \bm{\mu_i} ) + \sum_{i=1}^N  \sum_{k=1}^p h_{\lambda_1}^{(1)}(\beta_{ik}) + \sum_{l=1}^q h_{\lambda_2}^{(2)}(\alpha_l) ,
%\end{equation}
%where $\bm{\mu_i}(\bm{\beta_i}, \bm{\alpha})=\bm{X_i}\bm{\beta_i}+\bm{Z_i}\bm{\alpha}$,  $L(\cdot)$ is a loss function,   $h_{\lambda_1}^{(1)}(\cdot)$ and  $h_{\lambda_1}^{(2)}(\cdot)$  are  feature-selection penalties for individualized parameters and population-shared parameters  respectively, and $\lambda_1$, $\lambda_2$ are the corresponding tuning parameters.  
where $\parallel\cdot\parallel_2$ denotes  the Euclidean norm, and  $h_{\lambda_{N,m}}(\cdot)$  refers to a feature selection penalty function,  e.g.,  Lasso, adaptive Lasso, MCP or  SCAD.   Notice that the population-shared predictors $\bm{Z}_i$ mostly serve as control variables in applications,  and thus, in this paper,  we focus on  individualized variable selection of $\bm{\beta}_i$'s.

Next, we introduce some notations here.  Define  $\vec(\bm{b}_i)^N_{i=1}\equiv (\bm{b}_1^T, \ldots, \bm{b}_N^T)^T$ as a vectorization of a sequence of vectors $\{\bm{b}_i\}_{i=1,\ldots,N}$, and define $\bd(\bm{A}_i)_{i=1}^N \equiv \mbox{diag}(\bm{A_1}, \ldots, \bm{A_N})$ as a block-diagonal matrix with a sequence of matrices $\{\bm{A}_i\}_{i=1,\ldots,N}$ at the  diagonal.   We let  $\bm{\beta}_{(N)}=\vec(\bB_i)_{i=1}^N$ denote the  $Np$-by-$1$ grand vector of  individualized coefficients.  Furthermore, we denote $\bm{Y}=\vec(\bm{y}_i)_{i=1}^N$,  $\bm{X}=\bd(\bm{X}_i)_{i=1}^N$ and $\bm{Z}=\bd(\bm{Z}_i)_{i=1}^N$.   Without the penalty term in (\ref{objective:joint}), the ordinary least squares (OLS) estimator is obtained as 
\[
\setlength{\abovedisplayskip}{7pt}
\setlength{\belowdisplayskip}{5pt}
\vec(\hat{\bm{\beta}}_{(N)}^{OLS}, \hat{\bm{\alpha}}^{OLS})= [(\bm{X}, \bm{Z})^T(\bm{X}, \bm{Z})]^{-1}(\bm{X}, \bm{Z})^T\bm{Y},
\]
where the dimension of parameters ($Np+q$) will diverge as sample size $N$ increases. 
It is clear that the model in (\ref{objective:joint}) only utilizes individual-specific information in estimating the heterogenous coefficients $\bm{\beta}_i$'s,  which is hence named   individual-wise modeling.  As a result, this will lead to inefficient estimation and  over-fitting of a model, especially when the individual-specific information is limited, e.g., when the individual-wise measurement size $m$ is small.

\subsection{Multi-directional separation penalty}
To achieve more efficient estimation in  individualized modeling,  it is crucial and beneficial  to encourage grouping  some individuals which share similar treatment (covariates) effects. 
We propose a novel  penalization approach by providing multiple shrinking directions for individualized parameters and further utilizing  homogeneity information within  the  identified  subpopulations, which achieves simultaneous  parameter estimation, variable selection and individual subgrouping.

We consider a model which allows different subgroupings with respect to different heterogeneous-effect predictors. 
Specifically,  for the individualized coefficients $\bm{\beta}_{\cdot k}=(\beta_{1k}, \ldots,\beta_{Nk})^T$ of the $k$th heterogeneous-effect predictor ($k=1,\ldots,p$),  we assume that  there are $B_k$ subgroups as 
\begin{equation}\label{group}
\setlength{\abovedisplayskip}{7pt}
\setlength{\belowdisplayskip}{5pt}
  \beta_{ik} = \left\{ \begin{array}{cl}
  \bm\gamma_k^{(l)}, & \quad \text{if} \quad i \in \mathcal{G}_{k}^{(l)},  \quad l=1,\ldots, B_k-1\\
   0, &\quad \text{if} \quad i \in \mathcal{G}_{k}^{(0)} \end{array}\right., \quad \text{for} \; i=1,\ldots,N, 
\end{equation}
where each $\bm\gamma_k^{(l)}$ ($l=1,\ldots, B_k-1$) is an unknown non-zero sub-homogeneous effect  shared by individuals within the $l$th subgroup, and the index partition sets $\{\mathcal{G}_{k}^{(l)}\}_{l=0,1,\ldots,B_k-1}$  represent  the corresponding subgroup memberships in terms of the heterogeneous effects of  the $k$th predictor.   For ease of notation,  in the following,  we focus on the setting  where  there are two subgroups  with respect to each  heterogeneous-effect  covariate:  the  non-zero-effect group ($\beta_{ik}=\gamma_k, \; i \in \mG_k$) and the zero-effect group ($\beta_{ik}=0, \; i \in \mG_k^c$).  
%We denote  $\bm{\gamma}=(\gamma_1,\ldots,\gamma_p)'$ as the sub-homogeneous effect vector.  
%The extension to multiple subgroups is straightforward.

To achieve simultaneous variable selection and individual subgrouping,  we propose a penalized objective function with the sub-homogeneous effect $\bm{\gamma}=(\gamma_1, \ldots, \gamma_p)^T$ induced in a multi-directional separation penalty  (MDSP)   $s_{\lambda}(\cdot, \cdot)$  as 
%\begin{equation}\label{eq:Qi}
%\setlength{\abovedisplayskip}{5pt}
%\setlength{\belowdisplayskip}{7pt}
% Q^{ind}_{N,m}(\bm{\beta}_{(N)},\bm{\alpha}, \bm{\gamma})= \frac{1}{2} \sum_{i=1}^N  \parallel \bm{y_i}-  \bm{X_i}\bm{\beta_i} -\bm{Z_i}\bm{\alpha}\parallel_2^2+  \lambda_{N,m} \sum_{i=1}^N\sum_{k=1}^p s(\beta_{ik},\gamma_k),
%\end{equation}
\begin{align}
\setlength{\abovedisplayskip}{5pt}
\setlength{\belowdisplayskip}{7pt}
Q_{N,m}(\bm{\alpha}, \bm{\beta}_{(N)},\bm{\gamma})&=\frac{1}{2} \sum_{i=1}^N \big(\bm{y}_i-\bm{\mu_i}(\bm{\beta_i}, \bm{\alpha})\big)^T \bm{V_i}^{-1} \big(\bm{y}_i-\bm{\mu_i}(\bm{\beta_i}, \bm{\alpha})\big)+\sum_{i=1}^N\sum_{k=1}^p s_{\lambda}(\beta_{ik}, \gamma_k) \label{eq:Q} \\
&=L_{N,m}(\bm{\alpha}, \bm{\beta}_{(N)}) + S_{\lambda_{N,m}}(\bm{\beta}_{(N)},\bm{\gamma}),  \label{eq:LS}
\end{align}
where $\bm{\mu_i}(\bm{\beta_i}, \bm{\alpha})=\bm{X_i}\bm{\beta_i}+\bm{Z_i}\bm{\alpha}$.  To obtain more efficient estimation \citep{Liang:1986},  the within-individual  serial correlations are utilized by a weighting matrix $\bm{V}_i=\bm{A}_i^{\frac{1}{2}} \bm{R}_i\bm{A}_i^{\frac{1}{2}} $, where $\bm{A}_i$ is a diagonal matrix of marginal variance of $\bm{y}_{i}$ and $\bm{R}_i$ is a working correlation matrix.

The key component of the proposed model is the constructed multi-directional separation penalty (MDSP) function  $s_{\lambda}(\beta_{ik}, \gamma_k)$,  defined as 
\begin{equation}
\setlength{\abovedisplayskip}{5pt}
\setlength{\belowdisplayskip}{5pt}
 \label{separation}
s_{\lambda}(\beta_{ik}, \gamma_k)=\lambda_{N,m}\mbox{min}\big(|\beta_{ik}|, |\beta_{ik}-\gamma_k|\big),
\end{equation}
which is a piece-wise $L_1$-penalization function (Figure \ref{fig:penalty}), and $\lambda_{N,m}$ is a tuning parameter.  This multi-directional penalty  contains a double-summation, essentially  providing two  perspectives regarding  the proposed model in (\ref{eq:Q}).  First,  from an individual-wise  point of view,  the penalty term $\sum_{k=1}^p s_{\lambda}(\beta_{ik}, \gamma_k)$ applies on the $i$th individualized coefficients  $\bB_i=(\beta_{i1}, \ldots, \beta_{ip})^T$ given $\gamma_k$.    In contrast to the conventional  penalization approaches, the MDSP function $s_{\lambda}(\cdot, \gamma_k)$  provides each $\beta_{ik}$ ($k=1,\ldots, p$) an alternative shrinking direction  $\gamma_k$  in addition to zero, which essentially protects the strong signals from  being  pulled towards zero while shrinking those weak signals for sparsity pursuit, and also improves the variable selection accuracy. 
Although the underlying  sub-homogeneous effects $\gamma_k$'s are also unknown and to be estimated,  as illustrated in Figure \ref{fig:mdsp}, the proposed MDSP-estimator  reduces the bias on the non-zero coefficient estimators introduced by the traditional simultaneous  $L_1$-penalty,   as long as the estimated $\hat{\gamma}_k$ provides a roughly reasonable  direction  along  one dimension.

Indeed, the potential alternative direction provided by $\gamma_k$ can be estimated through  borrowing information from  other individuals who share similar effects, which is embedded in  
the other perspective of the proposed model in (\ref{eq:Q}).   From a population-wise view of point regarding the heterogeneous  effects of  the $k$th predictor,  the MDSP term  $\sum_{i=1}^N s_{\lambda}(\beta_{ik}, \gamma_k)$ turns to  group  the individualized  coefficients by separating the strong magnitude signals from the weak ones  that are close to zero,   which  roughly serves as a  center-based clustering analogous to the K-means approach.    Compared to pairwise grouping approaches such as the fusion penalty,  the MDSP model is more likely to ``separate'' the heterogeneous observations  given its construction,  rather than to ``combine'' them.  Moreover,  this center-based method also has less computational cost, with  $O(Np)$ penalty terms in contrast to the fusion-based clustering with $O(N^2p)$ penalty terms, which implies  a better computational scalability for a large sample size $N$.  In addition,  in the current model which performs  subgrouping on unobservable coefficients,   coefficients  estimation and  subgrouping are mutually influenced.   Therefore, the proposed method has an  advantage over the  two-stage procedure which carries out clustering  analysis based on  pre-estimated coefficients.

%%% reduce variance and bias
In addition,   the sub-homogeneous effects $\gamma_k$'s are estimated as the centers of the non-zero coefficient subgroups,  which significantly utilizes the  information from individuals in a homogeneous subpopulation and thus is more efficient than any single-individual-based estimation.    By pulling the individualized coefficients' estimators towards either zero or the $\hat{\gamma}_k$'s,  the MDSP model  reduces both the estimation bias and variance, and therefore gains extra accuracy in future prediction.    The above two-subgroup MDSP can be easily extended to  multiple subgroups,   even with additional constraints.  We illustrate the extension of three subgroups which allows  positive and negative effects of individualized  treatments  as 
\begin{equation*}\label{sp:tr}
\setlength{\abovedisplayskip}{5pt}
\setlength{\belowdisplayskip}{5pt}
s_{\lambda}(\beta_{ik}, \gamma_k^{+}, \gamma_k^{-})=\min\bigg(|\beta_{ik}|, |\beta_{ik}-\gamma_k^{+}|, |\beta_{ik}-\gamma_k^{-}|\bigg), \quad \mbox{s.t.} \quad  \gamma_k^{+} >0, \quad \gamma_k^{-}<0.
\end{equation*}
%which  shrinks heterogeneous coefficients either to zero, $\gamma_k^{+}$ (positive effect), or  $\gamma_k^{-}$ (negative  effect).

\subsection{Comparison with existing  subgroup analysis}
In this section, we make a few remarks comparing the proposed model with existing subgroup   models.  In addition to subgrouping on individualized regression coefficients, a key difference compared  to the most of the conventional subgrouping approaches \cite{Jacobs:1991,Gunter:2011,Pan:2013, Ma:2016, Zhu:2018},  is that our model in (\ref{eq:Q}) allows different subgroupings with respect to  heterogeneous coefficients of different predictors (\ref{group}). We refer to it as a \textit{covariate-specific subgrouping.}

Specifically,  we consider a simple example of a heterogeneous model with ten predictors:
\begin{equation}
\label{toy}
        \setlength{\abovedisplayskip}{5pt}
        \setlength{\belowdisplayskip}{5pt} 
        y_{i,t}=\beta_0+\beta_{i1}x_{i1,t}+ \cdots +\beta_{i10}x_{i10,t}+
        \varepsilon_{i,t}, \quad i=1,\ldots,N, t=1,\ldots,m,
\end{equation}
where each $\beta_{ik}$ ($i=1,\ldots,N,  k=1,\ldots,p$) is generated  independently from a Bernoulli distribution with a probability of 0.5.    Conventional clustering methods  target    subgrouping the coefficient vectors $\{\bm{\beta}_i \equiv (\beta_{i1},\ldots,\beta_{i10})^T\}$'s ($i=1,\ldots,N$), yielding subgroups corresponding to individuals sharing the same effects on all covariates.  As a result, this limits potential  applications,  as the inference is still at a population level, but not at an individual level.   For instance, if we further perform a variable selection based on the obtained subgroups, a variable will  be  selected/eliminated for all the individuals within the subgroup. 
 
 Furthermore,  population-level inference can also be unreliable  in many situations.  Consider the above example in (\ref{toy}).   The coefficient vector $\bm{\beta}_i$ essentially has $2^{10}=1,024$ unique  $(0,1)$ combinations leading to a potential $1,024$ underlying subpopulations.  However,  conventional clustering approaches are very likely to combine some of them as one group, e.g., $(1,\ldots,1,0)^T$ and $(1,\ldots,1,1)^T$ with finite samples,  which results in estimation biases.   Even under the assumption that all individuals are correctly classified into the true subpopulation,  the estimation for each  $\beta_{ik}$ is less efficient as it only utilizes approximately $N/1024$ samples in one subgroup, which  trades-off  small variance for unbiasedness.  In contrast, the proposed model with covariate-specific subgrouping is able to utilize almost $N/2$ samples in estimation of each parameter, which can achieve  unbiased and efficient estimation simultaneously,  while allowing each individual to have a unique coefficient vector.

\section{Theory}
\subsection{Double-divergence framework and notation}
%%%%%%%%%%%%%%%%%%%%%%%%%%%%%% Notations
In this section, we  lay out a new theoretical framework for individual-wise modeling inference and  population-wise subgrouping analysis in a double-divergence structure,  which allows     both sample size $N$ and individual measurements size $m$ go to infinity.   

We make contributions to two unique challenges under this framework.  First, % the individualized coefficients estimation and subgrouping are mutually influenced.  
as sample size $N$ increases, it is difficult to preserve the desired  strong oracle property of the individualized coefficients,  which enables each individual to utilize the true subpopulation information and thus to achieve  optimal estimation efficiency.  This is because  the number of individualized parameters is diverging and  a strong oracle property  essentially requires a subgrouping consistency,  that is,  classifying  all the individuals into the correct subpopulation.   We establish theoretical results indicating that the proposed estimator enjoys the strong oracle property  and we also outline the optimal divergence rates of $N$ with different assumptions.    Second,  in contrast to traditional longitudinal analysis, as the number of  individual measurements  $m$ increases, the individual-specific  correlation can have a significant effect on the  convergence rate of the estimator,  as the correlation matrices   in (\ref{eq:Q}) are  also  expanding.  We provide the convergence rate of the proposed estimator taking unknown correlation structure into account based on a double-divergence estimating equation.

We start by introducing some notation.   For a symmetric matrix $\bm{A}_{n \times n}$, let $\lambda_{min}(\bm{A})$ and $\lambda_{max}(\bm{A})$ be the smallest and the largest eigenvalues of $\bm{A}$, respectively.    For an arbitrary matrix $\bm{A}_{m\times n}(a_{ij})$, denote  $\|\bm{A}\|_2=\sqrt{\lambda_{max}(\bm{A}^T\bm{A})} $ as its $L_2$-norm, $\|\bm{A}\|_1=\max\limits_{1\leq j\leq n}(\sum_{i=1}^m |a_{ij}|)$ as its $L_1$-norm,  $\|\bm{A}\|_{\infty}=\max\limits_{1\leq i\leq m}(\sum_{j=1}^n|a_{ij}|)$ as its $L_{\infty}$-norm, and  denote $tr(\bm{A})$  as its trace.  For a vector $\bm{a}=(a_1,\ldots,a_n)^T$,  let $\|\bm{a}\|_0=\sum_{i=1}^n I_{\{a_i \neq 0\}}$.  Moreover, let $\bm{A} \circ \bm{B}$  denote the entrywise Hadamard product between two same-dimension matrices  and  let ``$\otimes$'' denote the Kronecker product.

 In addition,  we let $|\mathcal{G}_k|$  denote the cardinal norm of the index set $\mathcal{G}_k \subset \{i:1,\ldots,N\}$ where $\beta_{ik}=\gamma_k$ if $i \in \mathcal{G}_k$,  and $\mathcal{G}^c_k$ is its complement ($\beta_{ik}=0$).   We  denote  $\bm{\theta}=\vec(\bm{\beta}_{(N)}, \bm{\alpha})$ as a grand coefficients vector  and let   $\bm{\theta}^0=\vec(\bm{\beta}_{(N)}^0, \bm{\alpha}^0)$ be its true value,  and let $\bm{\gamma}^0$ be the true value of $\bm{\gamma}$.  Furthermore, we denote the true value of an individual coefficient $\bm{\beta}_i$ as $\bm{\beta_i^0}=\vec({\bm{\beta^0_{i,\mathcal{A}_i}}},\bm{0})$, where $\mathcal{A}_{i} \subset \{1,\ldots,p\} $ denotes the signal index sets such that $\bm{\beta}^0_{ik}=\bm{\gamma}^0_{k}$ if $k \in \mathcal{A}_i$.

%%%%%%%%%%%%%%%%%%%Double Divergence Estimating Equation

%We first provide a general view of the proposed framework.

%The proposed objective function (\ref{eq:Q}) consists of a loss function $L_{N,m}(\cdot)$ and a penalty function $S_{\lambda_{N,m}}(\cdot)$, where the  squared loss function   $L_{N,m}(\bm{\theta}) $ in (\ref{eq:LS})  can accommodate  %a double-divergence form along sample size diverging $N$ and $m$.  
The  individual-wise estimator without subgrouping refers to an unpenalized estimator minimizing   the squared loss function  $L_{N,m}(\bm{\theta}) $ in (\ref{eq:LS}),  which corresponds to solving the  quasi-likelihood estimating equation
 % Note that  it is equivalent to solve the following quasi-likelihood estimating equation
\begin{equation}\label{eq:G}
\setlength{\abovedisplayskip}{5pt}
\setlength{\belowdisplayskip}{7pt}
\bm{G}_{N,m}(\bm{\theta})=\sum_{i=1}^N  \bm{g}_i(\bm{\theta})=\sum_{i=1}^N \bm{U}_i(\bm{\theta})^T\bm{V}_i^{-1} \big(\bm{y}_i-\bm{\mu}_i(\bm{\theta})\big)=0,
\end{equation}
where  $\bm{U}_i(\bm{\theta})=\frac{\partial \bm{\mu}_i(\bm{\theta})}{\partial \bm{\theta}^T}$.   With  a  linear mean function, $\bm{U}_i(\bm{\theta})$  does not actually depend on unknown parameter $\bT$ and thus is suppressed as $\bm{U}_i$ for simple notation.   In addition,  we let
\setlength{\abovedisplayskip}{3pt}
\setlength{\belowdisplayskip}{5pt}
\begin{align*}
&\bm{D}_{N,m}=-\frac{\partial \bm{G}_{N,m}(\bm{\theta})}{\partial \bm{\theta}^T}=\sum_{i=1}^N \bm{U}_i^T\bm{V}_i^{-1}\bm{U}_i,\\
&\bm{H}_{N,m}= \mbox{Cov}(\bm{G}_{N,m}(\bm{\theta}))=\sum_{i=1}^N \bm{U}_i^T\bm{V}_i^{-1}\bm{\Sigma}_i\bm{V}_i^{-1}\bm{U}_i,
\end{align*}
where $\bm{\Sigma}_i=\mbox{Cov}(\bm{y}_i)=\bm{A}_i^{\frac{1}{2}} \bm{R}^0_i\bm{A}_i^{\frac{1}{2}}$ and $\bm{R}^0_i$ is the true correlation matrix.  
Note that $\bm{D}_{N,m}$ and $\bm{H}_{N,m}$  are  both $(Np+q)$-dimensional symmetric  matrices, which do not involve unknown parameter $\bm{\theta}$.  
Under the homogeneous variance assumption, $\bm{A}_i$ can be dropped.    In addition,  we usually assume $\bm{R}^0_i=\bm{R}^0$ and choose working correlation  $\bm{R}_i=\bm{R}$  for   $i=1, \ldots,N$. Due to the unknown true correlation $\bm{R}^0$, $\bm{D}_{N,m}$ and $\bm{H}_{N,m}$ are not necessarily equal, unless $\bm{R}$ is correctly specified.

Section A.2 of the Supplementary Materials lists a set of mild regularity conditions which are assumed in the following discussion. They are all standard assumptions made on regressors in penalized variable selection approaches and  longitudinal data models \citep{Xie:2003, Wang:2012},  with a small extension to the current individualized model setting.  In particular,  the standard assumptions  of $\bm{R}^0_i$  converging to  a constant positive definite matrix with eigenvalues bounded away from zero and infinity  \citep{Wang:2012}  might  not be  valid  here,  as  the dimension of $\bm{R}^0_i$ diverges  as the  individual measurement size $m$ diverges. We impose a mild regularity condition (A3) instead on the expanding correlation matrices which can be easily verified on a set of common correlation structures such as Exchangeable, AR-1 and Toeplitz.

%In general, for an estimator $\hat{\bm{\theta}}$ obtained by solving the estimating equation (\ref{eq:G}), under regularity conditions (A1)-(A2),  by Taylor's expansion, $(\hat{\bm{\theta}} - \bm{\theta}^0) = - \bm{D}_{N,m}^{-1} \bm{G}_{N,m}$, where we denote $\bm{G}^0_{N,m}=\bm{G}_{N,m}(\bm{\theta}^0)$  for ease of notation. 

%%%Individual-wise  unpenalized  estimator
\subsection{Oracle estimator and unpenalized individual-wise estimator}
In this section, we provide asymptotic results to the  individualized  estimator without penalization and the oracle estimator with true subgroup information.  Both of the two estimators play important roles in understanding the individual-wise model inference and in investigating the large sample property of the proposed MDSP estimator. 

The estimating equation $\bm{G}_{N,m}(\bm{\theta})$  contains double summations with   sample size $N$ and   individual measurement  size $m$ that both can diverge.  Therefore, the standard asymptotic results for $M$-estimators are not applicable here even with a fixed number of parameters  \citep{Xie:2003}.   The following lemma   implies that the consistency  of the unpenalized estimator $\hat{\bm{\theta}}^u$  solved from the equation $\bm{G}_{N,m}(\bm{\theta})=\bm{0}$ in (\ref{eq:G}) relies on the  information matrix $\bm{D}_{N,m}\bm{H}^{-1}_{N,m}\bm{D}_{N,m}$.

%%%%lemma 1
\begin{lemma}\label{lm:theta}
Under regularity conditions (A1)-(A-2) provided in the Supplementary Materials,   for any $\delta>0$, there exists a solution $\hat{\bm{\theta}}^u$ of the equation in (\ref{eq:G}) such that
\[
 P\bigg( p_{\bm{\theta}}^{-\frac{1}{2}} \|\bm{H}_{N,m}^{-\frac{1}{2}} \bm{D}_{N,m} (\hat{\bm{\theta}}^u - \bm{\theta}^0 ) \|_2 >\delta  \bigg) < \frac{1}{\delta^2},
\]
where  $p_{\bm{\theta}}=Np+q$ is the dimension of $\bm{\theta}$.   Moreover,  if  condition ($\mathcal{C}_a$):   $ \lambda_{min}(\bm{D}_{N,m}\bm{H}^{-1}_{N,m}\bm{D}_{N,m}) \rightarrow \infty$ holds, we have
\[
P\bigg(p_{\bm{\theta}}^{-\frac{1}{2}} \| \hat{\bm{\theta}}^u - \bm{\theta}^0 )\|_2 >\delta \bigg) \longrightarrow 0.
\]
\end{lemma}

\begin{remark}
\normalfont
The condition ($\mathcal{C}_a$) is a standard  condition  analogous to  the one in  \citep{Xie:2003} for  the weak consistency of a fixed-dimensional generalized estimating equation (GEE)  estimator.  In an independent  model where $\bm{R}^0=\bm{R}=\bm{I}_m$ or the working correlation $\bm{R}$ is correctly specified,   the information   $\bm{D}_{N,m}\bm{H}^{-1}_{N,m}\bm{D}_{N,m}$  reduces to $\bm{D}_{N,m}$. Notice that, in the individualized model setting, the divergence rate of the smallest eigenvalue of $\bD$ (the same as $\bH$) only depends on the number of individual measurements $m$.  Therefore, the condition ($\mathcal{C}_a$) essentially implies the divergence of $m$, that is, we need cumulative individual information to ensure  consistent estimation.  
\end{remark}
Lemma \ref{lm:theta} provides the consistency result under an $L_2$ norm (spectral norm),  which actually   requires a limited  sample size $N$, otherwise the parameter dimension $p_{\bT}$ will diverge as $N$ increases.  However, the proof of Lemma \ref{lm:theta} shows that,   as $m$ diverges, the consistency of $\hat{\bT}^u$ can be guaranteed  as long as  $N$ diverges with a limited rate.  We will have more discussion regarding this point later.

%under the spectral norm ($L_2$-norm).  For any fixed-dimensional estimator,  for example, the oracle estimator and the individual-wise estimator when $N$ is bounded,  the consistency in Lemma \ref{lm:theta}  is equivalent to $P\bigg(   \| \hat{\bm{\theta}} - \bm{\theta}^0 )\|_{\infty} >\delta \bigg) \rightarrow 0$.  However, if $p_{\bm{\theta}}$ is diverging,  we need additional conditions to ensure the stronger consistency under the $L_{\infty}$-norm.  More discussion will be provided later regarding  the proposed estimator when $N \rightarrow \infty$.

 %However, in contrast to \citep{Xie:2003}'s setting,   the proposed method results  in a diverging dimension of the information matrix % in the proposed model leads to which is  more  complicated.  In addition, to utilize subpopulation information,  the convergence rates for  estimators of different parameters  are of great importance and interest  in this paper. The following lemma provides a   convergence property for the estimating equation estimator from (\ref{eq:G}).

Next, we provide the theoretical results for the oracle estimator, which assumes being given the true  subpopulation information ($\mathcal{G}_k$,  $1 \leq k \leq p$) with respect to all individualized predictors.    This is equivalent to assuming that all individualized true signal sets $\mathcal{A}_i$'s ($ 1 \leq  i\leq N$)  are known.  Consequently, each  individualized oracle parameter  $\bm{\beta}^{or}_i$  is linked to the sub-homogeneous effect $\bm{\gamma}$ as   $\bm{\omega_i}\circ\bm{\gamma}=\bm{\beta}^{or}_i$ through an indicator vector  $\bm{\omega_i}=(\omega_{i1},\ldots,\omega_{ip})^T $,  where $\omega_{ik} = \mathbf{1}_{\{i \in \mathcal{G}_k\}}= \mathbf{1}_{\{k\in \mathcal{A}_i\}}$ and $\mathbf{1}_{\{\cdot\}}$ denotes an indicator function.   Hence there exists a mapping linking  two parameter spaces:  $\mathbf{R}^{p} (\bG)\mapsto \mathbf{R}^{Np} (\bB^{or}_{(N)}): \bm{\Omega}  \bm{\gamma}=\bm{\beta}^{or}_{(N)}$, where $\bm{\Omega}_{Np \times p} \equiv [\bm{\Omega}_1\;\cdots \; \bm{\Omega}_N]^T$ and $ \bm{\Omega}_i=\mbox{diag}(\bm{\omega}_i)$ is a diagonal matrix.
%We define $L_{N,m}^{or}(\bm{\alpha}, \bm{\gamma})=L_{Nm}(\bm{\alpha},\bm{\beta}_{(N)}( \bm{\gamma}))$.  
Therefore, by noting that $S_{\lambda_{N,m}}(\bm{\beta}^{or}_{(N)}, \bm{\gamma})=0$,   the oracle estimator is obtained as 
\begin{equation}
        \setlength{\abovedisplayskip}{5pt}
        \setlength{\belowdisplayskip}{5pt} 
\label{oracle}
 \vec( \hat{\bm{\gamma}}^{or} , \hat{\bm{\alpha}}^{or})=\argmin_{\bm{\alpha}, \bm{\gamma}} \sum_{i=1}^N \bigg(\bm{y}_i -\bm{X_{i}} (\bm{\omega}_i \circ \bm{\gamma}) - \bm{Z_{i}}\bm{\alpha}\bigg)^T \bm{V}_i^{-1}\bigg(\bm{y}_i -\bm{X_{i}} (\bm{\omega}_i \circ \bm{\gamma}) - \bm{Z_{i}}\bm{\alpha}\bigg), 
\end{equation}
and the oracle  individualized estimator is  $\hat{\bm{\beta}}^{or}_i=\bm{\omega}_i \circ \hat{\bm{\gamma}}^{or}$.  We first establish the asymptotic result for  the oracle estimator with an independent model to  reveal the  subpopulation effect on estimation.

% Theorem 1: independent oracle estimator
\begin{thm}\label{thm:or}
Under regularity conditions (A4)-(A6) provided in the Supplementary Materials,  suppose $\vec( \hat{\bm{\gamma}}^{or} , \hat{\bm{\alpha}}^{or})$ is the oracle estimator  of an independent model  obtained in (\ref{oracle}), where $\bm{R}^0=\bm{R}=\bm{I}_m$;   as  either  $m \rightarrow \infty$ or  $\min\limits_{1\leq k \leq p}(|\mathcal{G}_k|) \rightarrow \infty$, we have
\[
(\bm{H}^{or}_{N,m})^{\frac{1}{2}} \bigg(  \vec( \hat{\bm{\gamma}}^{or} , \hat{\bm{\alpha}}^{or})-  \vec( \bm{\gamma}^{0} , \bm{\alpha}^{0}) \bigg) \longrightarrow_d N\bigg(\bm{0}, \bm{I}_{p+q} \bigg),
\]
where $\bm{H}^{or}_{N,m} \asymp \bm{M}_{N,m}$,  and
 $\bm{M}_{N,m}=\mbox{diag}( \underbrace{N_1, \ldots, N_p}_p,  \underbrace{N_a, \ldots, N_a}_q)$ is a $(p+q)$-dimensional  diagonal matrix, in which,  $N_k= m|\mathcal{G}_k|$, $k=1,\ldots,p$,   and $N_a=mN$. The operator ``$\asymp$'' denotes that the matrix $\bm{H}^{or}_{N,m}$ has the same order as  $\bm{M}_{N,m}$.  The rigorous definition of ``$\asymp$'' and the explicit form of $\bm{H}^{or}_{N,m}$ are provided in Section A.4 of the Supplementary Materials.
\end{thm}

Theorem \ref{thm:or}  indicates that the convergence rates of the  oracle estimator  benefit from  increasing both $N$  and $m$, as it fully utilizes the subpopulation information and thus achieves  optimal  estimation efficiency.   In particular, the convergence rates of the sub-homogeneous-effect estimator $\hat{\gamma}_k$'s   are covariate-specific, corresponding to $\sqrt{N_k}$ ($1 \le k \le p$), respectively.  The asymptotic result for the oracle estimator with correlated data is further discussed in the next subsection.

\subsection{Multi-directional separation penalty estimator with correlated data}
In this section, we establish the large sample results for the proposed MDSP estimator with correlated data. In addition, we provide the optimal  divergence rate of $N$ that can be achieved while ensuring the oracle property of the proposed estimator.

Incorporating correlations on individual-wise measurements brings additional theoretical challenges to the double-divergence framework, as it involves  divergent-dimensional    correlation matrices  $\bm{R}_i$ and  $\bm{R}_i^0$. This makes it difficult to figure out the estimators' convergence rates.  In addition to condition ($\mathcal{C}_a$), we provide an alternative sufficient condition in the following theorem,  which could  simplify the verification and  discussion similar to {\citep{Xie:2003}.

%%Theorem 2 Correlated oracle estimator
\begin{thm}\label{cthm:or}
Let $\eta_{m}=\max\limits_{1\leq i \leq N} \{\lambda_{max} (\bm{R}_i^{-1}\bm{R}_i^0)\}$. Under regularity conditions (A3)-(A6) provided in the Supplementary Materials,  for the oracle estimator $\hat{\bm{\theta}}^{or}=\vec(\hat{\bm{\gamma}}^{or}, \hat{\bm{\alpha}}^{or})$ obtained in (\ref{oracle}),   we have
\[
\eta_{m}^{-\frac{1}{2}}\| (\bm{D}^{or}_{N,m})^{\frac{1}{2}} (\hat{\bm{\theta}}^{or}-\tilde{\bm{\theta}}^0)\|_2 \leq O_p(1), 
\]
where $\tilde{\bm{\theta}}^{0}=\vec(\bm{\gamma}^{0}, \bm{\alpha}^{0})$, and $\bD^{or}$ is the second-order derivative matrix for the  objective function in (\ref{oracle}). The explicit  form of $\bD^{or}$ is provided in Section A.4 of the Supplementary Materials;   Furthermore, if  condition $(\mathcal{C}^*_a)$:  $\eta_{m}^{-1}\lambda_{min}(\bm{D}^{or}_{N,m}) \rightarrow \infty$ holds,      then   $\hat{\bm{\theta}}^{or} \rightarrow_p \tilde{\bm{\theta}}^0$ under an $L_2$ norm.  
\end{thm}

Theorem \ref{cthm:or} indicates that the convergence  of  the estimator  depends on the divergence rate of $\eta_{m}$ and $\bm{D}^{or}_{N,m}$, where $\eta_{m}$ measures the  ``deviation'' between the working correlation structure $\bm{R}_i$ and the true correlation structure $\bm{R}_i^0$.   %The sufficiency of ($\mathcal{C}^*_a$) that implies  ($\mathcal{C}_a$)  is trivial by noting $\bm{H}_{N,m}  \leq \eta_{m}\bm{D}_{N,m}$. 
It is clear that if an appropriate working correlation matrix $\bm{R}_i$ is specified, we gain extra estimation efficiency by reducing $\eta_m$.  
However,  in general, as $m \rightarrow \infty$,  the value of $\eta_{m}$ is not always bounded.    Therefore, the convergence rate of the estimator could be slower than the optimal rate $\sqrt{m}$ and it may not converge to a normal distribution asymptotically  \citep{Xie:2003}.
We provide  more  discussion  with  a few  common cases and  some useful conditions in Section A.6 of the Supplementary Materials.

To finally establish the large sample theory for the MDSP estimator,  as well as providing the divergence rate of  sample size $N$,  we consider two sets of assumptions on random error $\bm{\varepsilon_{i}}$'s: \\
\indent ($\mathcal{I}_a$):  \hspace{2mm} Assume that $\bm{\varepsilon_{i}}=(\varepsilon_{i,1},\ldots,\varepsilon_{i,m})^T$ is independent and identically generated with mean zero and the  covariance matrix $\Sigma_{m}=\sigma^2 \bm{R}^0$, where $\sigma<\infty$, for $i=1,\ldots,N$; \\
\indent  ($\mathcal{I}_b$): \hspace{2mm} In addition to ($\mathcal{I}_a$), let $\bm{\varepsilon_{i}}^*= \bm{\Sigma}_m^{-\frac{1}{2}}\bm{\varepsilon_{i}}$, assuming that  $\bm{\varepsilon_{i}}^*$  is a sub-Gaussian vector,  that is,  $\mbox{P}(|\bm{a}^T\bm{\varepsilon_{i}}^*|>t) < 2\mbox{exp}(-\frac{t^2}{c_{\sigma}^2 \|\bm{a}\|_2^2})$ for any $\bm{a} \in \mathbf{R}^{m}$ and $t>0$, where $c_{\sigma}$ is a positive constant.  \\
In the independent-error model, the assumption in ($\mathcal{I}_b$) is equivalent to assuming marginal sub-Gaussian tails for $\varepsilon_{ij}$'s, which is a standard assumption in high-dimensional data models. Alternatively,  if the random errors are assumed to be normally distributed, then ($\mathcal{I}_b$) holds naturally for both independent and correlated data.

Based on the above conditions and results,  we establish the large sample  theory for the proposed estimator  under a double-divergence setting.

%Theorem 4 (Proposed estimator  oracle property)
\begin{thm}\label{cthm:beta}
Let  $\tau_{m}=\lambda_{min}(\bm{D}_{N,m} (\bm{H}_{N,m})^{-1} \bm{D}_{N,m})$.    Under regularity  conditions (A1)-(A6) provided in the Supplementary Materials,   suppose  $\frac{\lambda_{N,m}}{\tau_{m}} \rightarrow 0$ and  $\frac{\lambda_{N,m}}{\sqrt{\tau_{m}}} \rightarrow \infty$ holds,  there exists a local  minimizer $\vec(\hat{\bm{\alpha}}, \hat{\bm{\beta}}_{(N)},\hat{\bm{\gamma}})$ of the MDSP objective function  in (\ref{eq:Q});  as  $\tau_{m} \rightarrow \infty$,    we have
\[
\mbox{P}\bigg\{ \vec \big(\hat{\bm{\alpha}}, \hat{\bm{\beta}}_{(N)},\hat{\bm{\gamma}}\big) = \vec\big(\hat{\bm{\alpha}}^{or}, \hat{\bm{\beta}}^{or}_{(N)},\hat{\bm{\gamma}}^{or}\big) \bigg\} \longrightarrow 1,
\]
with \\
\indent (i) $N=o(\tau_{m})$,  if  Assumption ($\mathcal{I}_a$) holds, or  \\
\indent (ii) $\log(N)=o(\tau_{m})$,  if  Assumption ($\mathcal{I}_b$) holds. \\
The explicit forms of $\bD$ and $\bH$ are provided in Section A.7.2 of  the Supplementary Materials.  If the working correlation is correctly specified $\bm{R}_i=\bm{R}^0_i$, $1 \le i \le N$,   we have $\tau_{m}=\lambda_{min}(\bm{D}_{N,m})$.
\end{thm}

Theorem \ref{cthm:beta} indicates that the proposed estimator is the same as the oracle estimator,  which utilizes  most of the information  of the underlying subpopulation structure,    ensuring that the proposed estimator inherits  optimal  efficiency  from the oracle estimator and that the effects for each individualized predictor are correctly classified.   To summarize, we achieve both individual-wise variable selection consistency and covariate-wise subgroup identification consistency as follows.

%Corollary 3 (Model selection consistency)
\begin{corollary}[Uniform variable selection consistency]\label{cr:selection}
Under the same conditions  as in Theorem \ref{cthm:beta}, as $\tau_{m} \rightarrow \infty$, we have $P\bigg(\bigcap_{i=1}^N \{\hat{\mathcal{A}}_i=\mathcal{A}_i\}\bigg) \rightarrow 1$.
\end{corollary}

%Corollary 4 (Subgrouping consistency)
\begin{corollary}[Uniform subgroup identification consistency]\label{cr:subgrop}
Under the same conditions  as in Theorem \ref{cthm:beta}, as $\tau_{m} \rightarrow \infty$, we have $P\bigg(\bigcap_{k=1}^p \{\hat{\mathcal{G}}_k=\mathcal{G}_k\}\bigg) \rightarrow 1$.
\end{corollary}

Theorem \ref{cthm:beta} also provides the optimal  divergence rates of $N$, which depends on the order of $\tau_m$,  to ensure the oracle property for the proposed estimator given different assumptions on random errors.  
It is apparent that  $\tau_{m} \rightarrow \infty$ as $m\rightarrow\infty$, while the explicit order of $\tau_{m}$ is not easy to obtain in general  as it involves  unknown divergent-dimension correlation structures.  Under additional  assumptions or given specific structures  on the correlation matrices,  we are able to establish it  as discussed in Section A.6 of the Supplementary Materials.  In particular, with an independent error-model,  by noting $\tau_{m}=m$, we have a simplified result as stated in the following corollary.

%Corollary 4 (Independent estimator oracle property)
\begin{corollary}[Oracle property in independent model]
\label{ccr:beta}
Under the same conditions as in Theorem \ref{cthm:beta},  suppose $\bm{R}_i=\bm{R}_i^0=\bm{I}_m$, for $1 \le i \le N$,  if  $\frac{\lambda_{N,m}}{m} \rightarrow 0$ and  $\frac{\lambda_{N,m}}{\sqrt{m}} \rightarrow \infty$, there exists a local  minimizer $\vec(\hat{\bm{\alpha}}, \hat{\bm{\beta}}_{(N)},\hat{\bm{\gamma}})$ of the MDSP objective function  in (\ref{eq:Q});  as  $m \rightarrow \infty$,    we have
\[
\mbox{P}\bigg\{ \vec \big(\hat{\bm{\alpha}}, \hat{\bm{\beta}}_{(N)},\hat{\bm{\gamma}}\big) = \vec\big(\hat{\bm{\alpha}}^{or}, \hat{\bm{\beta}}^{or}_{(N)},\hat{\bm{\gamma}}^{or}\big) \bigg\} \longrightarrow 1,
\]
with  (i) $N=o(m)$  if  Assumption ($\mathcal{I}_a$) holds,  or (ii) $\log(N)=o(m)$  if  Assumption ($\mathcal{I}_b$) holds.
\end{corollary}

Lastly, we consider applying the MDSP model to a new dataset such as a new individual  which is usually challenging but also crucial for subgroup analysis.   Since this framework focuses on  unobservable predictor effects, we assume to have a semi-new individual which has initial observations $\bm{y}^*_{i}$ with independent errors.  Given a   pre-estimated  sub-homogeneous effect $\hat{\bG}=(\hat{\gamma}_1, \ldots, \hat{\gamma}_p)^T$ from a training dataset, we fit the model on a semi-new individual as 
\begin{equation}\label{objective:idv}
 Q_{i,m^*}(\bm{\beta^*_i,\bm{\alpha}^*} | \hat{\bm{\gamma}})=\frac{1}{2} \parallel \bm{y}^*_i- \bm{X}^*_{i} \bm{\beta^*_i}-\bm{Z^*_{i}} \bm{\alpha}^* \parallel_2^2    +     (\lambda_{m^*})\sum_{k=1}^p s(\beta^*_{ik},\hat{\gamma}_k).
\end{equation}

%%% Theorem 4 Semi-new individual
\begin{thm} 
\label{thm:new}
Suppose $\sqrt{m^*}(\hat{\bm{\gamma}}-\bm{\gamma}^{0}) \le O_p(1)$.  Under regularity conditions (A1)-(A6)  provided in  the Supplementary Materials,   there exists a minimizer $\hat{\bm{\beta}}^*_i=\vec({\hat{\bm{\beta}}^*_{i,\mathcal{A}_i}},{\hat{\bm{\beta}}^*_{i,\mathcal{A}_i^c}})$ of (\ref{objective:idv}),   if $\lambda_{m^*} \rightarrow 0$ and $\lambda_{m^*}/\sqrt{m^*} \rightarrow \infty$, as $m^* \rightarrow \infty$, we have  
\[
 \mbox{P}(\hat{\bm{\beta}}^*_{i,\mathcal{A}^c_{i}} =0) \rightarrow 1 \quad \text{and} \quad \mbox{P}(\hat{\bm{\beta}}^*_{i,\mathcal{A}_{i}} =\hat{\bm{\gamma}}_{\mathcal{A}_{i}}) \rightarrow 1,
 \]
 where $\mathcal{A}_i$ denotes the true signal index set for the $i$th semi-new individual. 
 \end{thm}

Theorem  \ref{thm:new} provides an insight from an individual-wise perspective about how the MDSP enhances  individualized model inference on variable selection and model estimation. 
As a given $\hat{\bG}$ provides a reasonably good direction towards  sub-homogeneous effects, the individualized  estimator for the semi-new individual is able to achieve selection consistency even with a limited number of observations. The theorem does not require that the given estimator $\hat{\bG}$ is more efficient than the individualized  estimator which is  based on new observations only (with an order of $\sqrt{m^*}$).   However,  if $\hat{\bG}$ is obtained from a larger training sample with a convergence rate beyond $\sqrt{m^*}$, a single-individual based model can achieve a faster convergence rate  inherited  from the  given $\hat{\bG}$.

The  proofs of  all  of the theoretical results  are provided in  Appendix  A of  the Supplementary Materials.

\section{Computation}
\subsection{ADMM Algorithm}
The optimization problem  of the objective function  in (\ref{eq:Q})  is  challenging as it  involves the  non-convex penalty function  with an unknown sub-homogeneous-effect parameter, yielding non-separable parameters in estimation.  To achieve computational scalability, we propose an efficient ADMM-based algorithm \cite{Boyd:2011}, which decomposes  the original optimization into several smaller pieces that can be solved more easily.

To minimize the objective function  in (\ref{eq:Q}), we introduce a set of constraints $\beta_{ij}=\nu_{ij}$, $1\le i \le N$, $1 \le j \le p$, and consider a new constraint optimization problem 
\begin{equation}
\label{eq:copt}
\setlength{\abovedisplayskip}{5pt}
\setlength{\belowdisplayskip}{7pt}
\min_{\bA,\bB,\bN,\bG} L_{N,m} (\bm{\alpha}, \bm{\beta}) + S_{\lambda_{N,m}}(\bm{\nu},\bm{\gamma}), \quad s.t. \quad \bB=\bN,
\end{equation}
where $\bm{\beta}_{Np \times 1}\equiv(\beta_{ij})_{1\le i \le N, 1 \le j \le p}$ and $\bm{\nu}_{Np \times 1}\equiv(\nu_{ij})_{1\le i \le N, 1 \le j \le p}$. To solve (\ref{eq:copt}), we take the ADMM algorithm with the augmented Lagrangian function as
\begin{equation}
\label{eq:admm}
\setlength{\abovedisplayskip}{5pt}
\setlength{\belowdisplayskip}{7pt}
\cL(\bA, \bB, \bN, \bG)=L_{N,m}(\bm{\alpha}, \bm{\beta}) +S_{\lambda_{N,m}}(\bm{\nu},\bm{\gamma}) + \bm{\Lambda}^T(\bm{\beta}-\bm{\nu})+\frac{\kappa}{2} \| \bm{\beta}-\bm{\nu}\|_2^2,
\end{equation}
where $\bm{\Lambda}_{Np\times 1}\equiv (\Lambda_{ij})_{1\le i \le N, 1 \le j \le p}$ is the  Lagrangian multiplier,  and $\kappa$ is a fixed augmented parameter.  We update $\{\bA, \bB\}$, $\{\bN, \bG\}$ and $\bm{\Lambda}$ alternately at the $(l+1)$th iteration as follows:
\begin{align}
\{\bA^{(l+1)}, \bB^{(l+1)}\} &=\argmin_{\bA, \bB} L_{N,m} (\bm{\alpha}, \bm{\beta}) + \frac{\kappa}{2} \| \bB-\bN^{(l)}+\kappa^{-1}\bm{\Lambda}^{(l)}\|_2^2, \label{admm1}\\
\{\bN^{(l+1)}, \bG^{(l+1)}\} &=\argmin_{\bN, \bG} S_{\lambda_{N,m}}(\bm{\nu},\bm{\gamma})  + \frac{\kappa}{2} \| \bB^{(l+1)}-\bN+\kappa^{-1}\bm{\Lambda}^{(l)} \|_2^2, \label{admm2}\\
\bL^{(l+1)} &=\bL^{(l)}+\kappa(\bB^{(l+1)}-\bN^{(l+1)}). \nonumber
\end{align}

The  optimization  in (\ref{admm1})  turns to be  a quadratic  minimization problem given a  specified working correlation structure, which leads to an explicit solution.  We recommend a one-step moment estimation for the correlation structure $\bm{R}_i$  using  the individual-wise estimator from an independent model.  The objective function in the second optimization can be split into $p$ parallel pieces based on different heterogeneous covariates as
\begin{equation}
\label{admm22}
\setlength{\abovedisplayskip}{5pt}
\setlength{\belowdisplayskip}{7pt}
\argmin_{ \bN_{\cdot j} } \; \sum_{i=1}^N \bigg\{ \frac{\kappa}{2}(\nu_{ij} - \beta_{ij}^{(l+1)}-\kappa^{-1}\Lambda_{ij}^{(l)})^2 + \lambda_{N,m}   \min(|\nu_{ij}|, |\nu_{ij}- \gamma_j|) \bigg\},
\end{equation}
for $j=1,\ldots,p$, where $\bN_{\cdot j}=(\nu_{1j},\ldots,\nu_{Nj})'$.  Along the $j$th heterogeneous covariate, we iteratively  estimate $\bN_{\cdot j}$ and $\gamma_j$ with fixed $\bB^{(l+1)}$ and $\bL^{(l)}$. 
Specifically, given $\gamma_j$, the $\nu_{ij}$'s ($i=1,\ldots,N$) in  (\ref{admm22}) can be estimated separately with explicit solutions, and given $\nu_{ij}$'s, the $\gamma_j$ can be estimated via a one-dimensional exhaustive grid-search. Since all those pieces only involve   univariate optimization,  the minimization of (\ref{admm22}) can be solved easily.  More implementation details  in (\ref{admm1}),  (\ref{admm2}) and (\ref{admm22}) are provided in Section B.4 of  the Supplementary Materials.  The proposed algorithm is outlined in Algorithm \ref{alg}.

\begin{algorithm}
\caption{ADMM algorithm with parallel computing}
\label{alg}
 \textbf{Initialization}. Initialize $\bN^{(0)}, \bG^{(0)}$. Set $\lambda_{N,m}$ and $\kappa$.  Set $\bL=\bm{0}$. Set stopping tolerance levels $\epsilon_1$ and $\epsilon_2$.\\
 \indent For $l=0,1,2,\ldots$\\
 \textbf{Step 2}. Update $\{\bA^{(l+1)}, \bB^{(l+1)}\}$ via (\ref{admm1}).\\
 \textbf{Step 3}. Update $\{\bN^{(l+1)}_{\cdot j}, \gamma_j^{(l+1)}\}$ via (\ref{admm22}) with parallel computing over $j=1,\ldots,p$.\\
 \textbf{Step 4}. Update $\bL^{(l+1)} =\bL^{(l)}+\kappa(\bB^{(l+1)}-\bN^{(l+1)})$.\\
 \textbf{Step 5}. (Stopping Criterion) Iterate Steps 2-4 until $\big \{\parallel \bB^{(l+1)}- \bB^{(l)} \parallel_{2}/(Np)  + \parallel \bA^{(l+1)} - \bA^{(l)} \parallel_{2}/q+\parallel \bG^{(l+1)} - \bG^{(l)} \parallel_{2}/p \big\}< \epsilon_1$ and  $\parallel\bm{r}^{(l+1)}- \bm{r}^{(l)} \parallel_{2} < \epsilon_2$, where $\bm{r}^{(l)}=\bB^{(l)}-\bN^{(l)}$.
\end{algorithm}

%convergence theorem
 \begin{prop}\label{thm:convergence}
For the objective function in (\ref{eq:Q}), with a sufficiently large $\kappa$, the estimator  sequence generated by the proposed ADMM Algorithm \ref{alg}  converges to a stationary point of  (\ref{eq:Q}) subsequently.
\end{prop}

The proof of Proposition \ref{thm:convergence} can be shown by verifying the conditions R1-R3 in Proposition 1 of \cite{Zhu:2019}.  In practice, the iterative estimators may converge to a local minimizer due to the non-convex objective function.   Multiple initial values can be applied  to identify the optimum value.    In fact,   most   individuals are not sensitive to initial values except  the ones close to the boundaries of subgroups. Heuristically, if $\lambda_{N,m}/\gamma_k$ is small,  implying  that the true effects $\bm{\gamma}$ are strong, then the coefficient estimators  are likely consistent.    Therefore, we recommend  using  a warm-start for initialization,  which can be obtained by using the individual-wise least square estimator or the proposed MDSP estimator with a very small value of  $\lambda_{N,m}$ and a random initialization.

% In addition, we recommend a step-wise tuning in practice, that is,  we initialize  the tuning parameter by a  very small value  and increase it to the specified value  as the number of iterations increases.

\subsection{Tuning and subgroup number selection}

In this paper, we  tune  the shrinkage parameter $\lambda_{N,m}$  based on  the generalized cross-validation (GCV) method as suggested by \citep{Pan:2013},   which can be regarded as an approximation of leave-one-out cross-validation.  Specifically,  the GCV is defined as
\begin{equation*}\label{gcv}
\setlength{\abovedisplayskip}{5pt}
\setlength{\belowdisplayskip}{7pt}
GCV(\mbox{df}) =  \frac{RSS}{(mN-\mbox{df})^2}= \frac{\|\bm{Y} -\hat{\bm{Y}}\|_2^2}{(mN-\mbox{df})^2},
\end{equation*}
where $\mbox{df}$ is the degree of freedom used in estimating the $\hat{\bm{Y}}$.   In this  setting, the degree of freedom  cannot  simply be  treated   as the total number of non-zero parameters, since some of the coefficient estimator $\hat{\beta}_{ik}$'s are shrunk to the exact  sub-homogeneous effect $\hat{\gamma}_k$.  \citep{Pan:2013} suggests a generalized degree  of freedom (GDF), however,  which is computationally costly.  Approximately, here we define the degree of freedom ($\mbox{df}$) as  the total number of  unique non-zero coefficient estimators,  and  the tuning parameter $\lambda_{N,m}$ is thus selected  by a grid-based search to minimize the GCV.

In general, the proposed method allows a multi-subgroup setting  as defined in (\ref{group}),  while the number of subgroups is usually unknown and its selection is always challenging.   In practice, we could specify the  subgroup numbers  according to  known scientific  information or a  particular target such as  exploring the positive and negative treatment effects.   Alternatively,   we can select the number of subgroups based on  a data-driven approach.     One option is to  adopt the idea of the jump statistic \citep{Sugar:2003}  or  the gap statistic \citep{Tibshirani:2001}  based on the warm-start estimators. %, e.g.,  the individual-wise least squares estimator or the MDSP estimator with a very small value of  $\lambda_{N,m}$.   %This is easy to implement but might not be reliable, as in  the two-step procedure, the pre-estimators are treated as observed responses  which do not  change as the  number of subgroups changes.
In addition, \cite{Ma:2016}  provides a subgroup number selection strategy  based on  the modified Bayesian Information Criterion \citep{Wang:2007}. Specifically,   for the $k$th predictor, the number of subgroups $B_k$ is selected  by minimizing
\begin{equation*}\label{eq:BIC}
\setlength{\abovedisplayskip}{5pt}
\setlength{\belowdisplayskip}{7pt}
\mbox{BIC}(B_k) = \log\bigg( \sum_{i=1}^N\sum_{t=1}^m \{y_{i,t}-\hat{\mu}_{i,t}(B_k)\}^2 /(mN)\bigg)+b_{N,m} \frac{\log(mN)}{mN} (B_k+q-1),
\end{equation*}
 where $b_{N,m}$ is a positive number depending  on $N$ and $m$.    When $b_{N,m} = 1$, the modified BIC reduces to the traditional BIC  \citep{Schwarz:1978}.  For the high-dimensional setting,  we follow \citep{Wang:2009}  to take  $b_{N,m} = c\log(\log(p_{\bm{\theta}}))$, where  $p_{\bm{\theta}}=Np+q$ and $c=2$.   To extend  to multivariate  individualized predictors,   we  select  the number of subgroups for one predictor  while  fixing other individualized coefficients with  individual-wise least squares estimators.

\section{Numerical Study}
\subsection{Individualized Regression and Model Robustness}
In this section, we provide simulation studies to investigate the numerical performance of the proposed method in finite samples.  In the first simulation study,  we consider a heterogeneous regression model with  two population-shared variables and one individualized variable which, for example,  can be an interested treatment effect: 
\begin{equation}
\label{eq:sim1a}
\setlength{\abovedisplayskip}{7pt}
\setlength{\belowdisplayskip}{10pt}
y_{i,t}=\alpha_0+ \alpha_1z_{i1,t}+ \alpha_2z_{i2,t} + {\beta_{i}}x_{i,t}+\varepsilon_{i,t}, \quad i=1,\ldots,N,  \quad t=1,\ldots,m.
\end{equation}
We set the  sample size $N = 40, 100$, and the individual measurement size $m=10, 20$.  The  individualized coefficients are set as $\bm{\beta}=(\beta_{1},\ldots, \beta_{N})'=(\underbrace{\gamma,\ldots,\gamma}_{N/2},\underbrace{0,\ldots,0}_{N/2})'$,
where $\gamma$ is the true sub-homogeneous effect  chosen as 1 or 2,  and  the population parameters are $\alpha_0 = \alpha_1 =\alpha_2=1$. The covariates $z_{i1,t}$, $z_{i2,t}$ and $x_{i,t}$ are generated from $N(0,1)$.  The random error $\varepsilon_{i,t}$'s are independently generated  from $N(0,1)$.

We compare the performance of the proposed  model (MDSP) with five regularized variable selection approaches, namely, the Lasso \citep{Tibshirani:1996} implemented by R package \emph{glmnet} (version 2.0-2) \citep{Friedman:2010},  the adaptive Lasso (AdapL) \citep{Zou:2006} solved by  R package \emph{parcor} (version 0.2-6) \citep{Kraemer:2009},  the SCAD \citep{Fan:2001}  and the MCP \citep{Zhang:2010}  implemented by R package \emph{ncvreg} (version 3.5-1) \citep{Breheny:2011}, and the fused Lasso (FusedL) \citep{Tibshirani:2005} solved by R package \emph{penalized} (version 0.9-50) \citep{Goeman:2017}.  Note that there are $N+3$ variables and $Nm$ observations for the above five conventional regularization models.  %For the fused Lasso, we order estimators of  the individualized coefficients based on the least squares estimation as the fused Lasso only imposes $L_1$-penalties on adjacent coefficients.
In addition, we also compare  two non-variable-selection models, namely, the individual-wise model (Sub) obtaining individualized least-square estimators, and the homogeneous model (Homo) assuming  $\beta_i=\beta_h$, for $i=1,\dots,N$.  
To evaluate the performance of these  approaches on  individual variable selection and prediction, we calculate the correct variable identification rate (CVSR: rate of correctly identifying $\beta_{i}$'s to be either zero or non-zero),  sensitivity (true positive rate: $P(\hat{\beta}_i \neq 0 |\beta_i \neq 0)$) and  specificity (true negative rate: $P(\hat{\beta}_i=0 |\beta_i=0)$),  and the root mean square error (RMSE): $\| \hat{\bm{\beta}}- \bm{\beta}\|_2$, where $\bm{\beta}=(\beta_{i1},\ldots,\beta_{iN} )'$ are the true values of  coefficients.

Table \ref{table:s1armse} provides the average of root mean square errors (RMSE) based on 100 simulations while Figures \ref{fig:s1armse1} and \ref{fig:s1armse2} are the boxplots of the RMSE for all approaches.  The proposed method  has the smallest RMSE in all settings,  which has an improvement  of  at least  $20\%$ ($m=10$) and  $71\%$ ($m=20$) compared to  other methods for both sample sizes $N=40,100$ when $\gamma=1$.  The improvement is more significant reaching   $150\%$ ($m=10$) and  $250\%$ ($m=20$) when subgroups are separated well ($\gamma=2$).  This is because  the proposed method is able to borrow strength from different individuals within the same subgroup in estimating individualized coefficients. In addition,  Figures \ref{fig:s1avs2} and \ref{fig:s1avs4} provide the boxplots of CVSR, sensitivity and specificity ($N=100$) for all of the variable selection approaches. The proposed method (MDSP) clearly  outperforms the other conventional penalization approaches in terms of the highest CVSR and the specificity rates.  Additional tables and boxplots summarizing the estimation of sub-homogeneous effects,  CVSR,  sensitivity and   specificity are provided in Section B.3 of  the Supplementary Materials.

%Especially when the subgroup homogeneous effect is large ($\gamma=2$).  Although all models achieve  similar rates on sensitivity, the proposed model leads much higher specificity rates.  

%%%%%%%%%%%%%%%%%%Roubustness%%%%%%%%%%%%%%%%%%%%%%%
In unsupervised subgrouping analysis, determining the number of subgroups is  always  challenging.  Here we adopt the modified-BIC-based strategy  introduced in Section 4.2.  In the interest of space,   an additional simulation study investigating the selection of subgroup numbers is reported in Section B.1 of the Supplementary Materials.

Next we test the robustness of the proposed model when the number of subgroups is misspecified.  We generate the data as in model (\ref{eq:sim1a}) under two scenarios: one has a population homogeneous predictor ($\beta_i=\gamma=2, i=1,\ldots,N$) and the other generates  individualized  coefficients  with three subgroups ($\gamma_0=0, \gamma_1=-3, , \gamma_2=1$)  with balanced size.  %We set the sample size $N=60$ and the individual measurement  size $m=10$.  
For both  scenarios,  we fit the proposed model assuming two subgroups ($\beta_i=0,\gamma$).

Table  \ref{table:s2misp} provides the average RMSEs and CVSRs for the proposed method,  the individual-wise model and the five other regularized methods described in Section 5.1.  Figure  \ref{fig:missp}   illustrates the estimation of individualized coefficients from the proposed model.  In general, the proposed method is robust against the misspecification of subgroup numbers in terms of the consistently  smallest RMSE and the highest CVSR among all methods.  
Specifically,  the MDSP model does not suffer from the homogeneous-effect setting,  as all individuals   are essentially shrunk towards a  unique non-zero group effect.  
In the  scenario with three true subgroups,   the subgroup with a relatively stronger signal ($\gamma_1=-3$)  is successfully identified which gains more estimation efficiency, while the subgroup with the weaker effect ($\gamma_2=1$) is shrunk towards zero which does not have extra loss as  it   is  just  equivalent to the Lasso estimator.

 %In the case of homogeneous effect, all models perform similarly in selecting the true variable for all individuals. However, the proposed method has the smallest RMSE among all methods with  a $170\%$ reduction. In addition, in the case when there are fewer assumed subgroups than is true, the proposed method still has the best correct variable selection rate,  and reduces the RMSE at least $14\%$ compared to the other methods.

\subsection{Correlated data and application on semi-new individual}
In this subsection, we investigate the performance of the proposed model  utilizing within-individual correlation and  its application on newly observed individuals. 
We consider an individual-wise model of two individualized predictors with serial correlations:
\begin{equation}
\label{eq:sim2}
\setlength{\abovedisplayskip}{5pt}
\setlength{\belowdisplayskip}{10pt}
y_{i,t}=\alpha_0+ \alpha_1z_{i1, t}+ \alpha_2z_{i2, t} +{\beta_{i1}}x_{i1, t} + {\beta_{i2}}x_{i2, t} +\varepsilon_{i,t}, \quad i=1,\ldots,N,  \quad t=1,\ldots,m.
\end{equation}
The  individualized coefficients $\bm{\beta}_1=(\beta_{11},\ldots, \beta_{N1})^T$ and  $\bm{\beta}_2=(\beta_{12},\ldots, \beta_{N2})^T$ are generated as 
\begin{equation*}
\setlength{\abovedisplayskip}{7pt}
\setlength{\belowdisplayskip}{10pt}
  \bm{\beta}_1=(\underbrace{\gamma_1,\ldots,\gamma_1}_{N/2},\underbrace{0,\ldots,0}_{N/2}), \qquad \bm{\beta}_2=(\underbrace{0,\ldots,0}_{N/2},\underbrace{\gamma_2,\ldots,\gamma_2}_{N/2}),
\end{equation*}
where $\gamma_1=1$ and $\gamma_2=-2$.  %We choose the  sample size $N = 20, 80$ and the cluster size $m= 10, 20$.  
The covariates $z_{i1, t}$, $z_{i2, t}$, $x_{i1, t}$ and $x_{i2, t}$ are generated from $N(0,1)$.  The random error $\bm{\varepsilon_{i}}=(\varepsilon_{i,1},\ldots,\varepsilon_{i,m})^T$ is    generated from a multivariate normal distribution with mean $\bm{0}$ and covariance $\sigma^2\bm{R}(\rho)$, where $\bm{R}(\rho)$ is the correlation matrix which has either an AR-1 or exchangeable structure with $\sigma=1$ and $\rho=0.5$.

Table \ref{table:s1brmse} summarizes the average RMSEs of the MDSP model using different working correlation structures  compared  to the independent model.  In general,  the proposed model utilizing within-individual correlation information achieves smaller RMSE than the independent model. In particular, if the correct working structure is  specified, the RMSE can be reduced  at least $40\%$  compared to the one obtained using independent structure.

As an unsupervised learning, subgrouping analysis has a great challenge in dealing with the new individuals unless additional assumptions are imposed,   as in   subgroup membership  depending  on some other observable variables. However, these assumptions are essentially difficult to validate in practice.  Since this paper targets  non-observable covariates effects, following the existing literature about individualized dosage \cite{Zhu:2016, Diaz:2012},  here we consider a semi-new individual with a  limited number of initial individual observations. Specifically,  we generate a semi-new individual   with $m^*$  initial observations $\bm{y}^*_i=(y^*_{i1}, \ldots, y^*_{im^*})^T$ with covariates $\bm{x}_{ik}^*$'s and $\bm{z}_{ik}^*$'s ($k=1,2$) following  (\ref{eq:sim2}), for $i=1,\ldots,N^*$,  with independent errors,  where the coefficients $\beta^*_{i1}$ and $\beta^*_{i2}$ are generated from a Bernoulli distribution with a  probability of $0.5$.   We first estimate the sub-homogenous effects $\tilde{\gamma}_1$ and $\tilde{\gamma}_2$  by fitting  an MDSP model on a training set of $100$ individuals,  each individual with $20$ individual measurements.  For the $i$th semi-new individual, we apply the MDSP model given $(\tilde{\gamma}_1,\tilde{\gamma}_2)$: 
\[
\setlength{\abovedisplayskip}{5pt}
\setlength{\belowdisplayskip}{10pt}
\min_{\bm{\alpha}^*, \beta^*_{i1},\beta^*_{i2}} \| \bm{y}_i^* - \alpha^*_0 -  \alpha^*_1\bm{z}_{i1}^* -  \alpha^*_2\bm{z}_{i2}^*-  \beta_{i1}\bm{x}_{i1}^*-  \beta_{i2} \bm{x}_{i2}^*\|_2^2 + \lambda_s \sum_{k=1}^2  s(\beta^*_{ik}, \tilde{\gamma}_k).
\]

We investigate the parameter estimation (RMSE) and the variable selection (for $\beta_1$ and $\beta_2$)  on a semi-new individual using the MDSP model,  the individual-specific linear model,  and the individual-specific Lasso model.  For the linear model, the variable selection is based on the marginal p-value with a significance level of 0.05.  All results are evaluated based on  $N^*=100$  semi-new individuals with $m^*$ varying  from $6$ to $20$.  We  add a homogeneous model  estimator from the training as a reference. 

Figure \ref{fig:newsub} shows that the MDSP model consistently achieves the smallest RMSE values,  indicating the most efficient prediction accuracy,  and also has the best  accuracy in predictor selection/elimination. The improvement  of the MDSP model is more significant as the  semi-new individual has fewer initial observations,  e.g., when $m^*=6$, the MDSP model reduces the RMSE value by 476\% and 62\% compared  to the OLS model and the Lasso model, respectively. In addition, the MDSP model also consistently outperforms the homogeneous model  with an improvement of at least 34\%  (and up to 250\% as $m$ increases)  in the RMSE value.

\section{Real Data Application}

In this section, we apply the proposed individualized variable selection method  to the Detroit Neighborhood Health Study (DNHS) ({\ttfamily https://dnhs.unc.edu/}), which is a representative longitudinal study investigating genetic variation or traumatic events effects on mental disorders of African American adults in Detroit, Michigan.

The DNHS contains blood samples and five-wave surveys which ask questions about demographics, traumas, stressful events, and post-traumatic stress disorder (PTSD). The survey at each wave includes a post-traumatic checklist (PCL) based on incident trauma exposures, which is a $17$-item self-reported measure of PTSD symptoms. We treat the average of  $17$ PCL scores as the response variable with a logarithm transformation. Studies \cite{Rusiecki:2013, Chen:2016} show that pathophysiology of PTSD is associated with DNA methylation (DNAm) in glucocorticoid receptor regulatory network (GRRN) genes, since the process is intrinsically linked to gene regulation. 
To identify cytosine-phosphate-guanine (CpG) sites in GRRN genes which are significantly associated with PTSD,  we use DNAm values at $1648$ CpG sites as potential predictors. 

Specifically, we target  investigating the potential heterogeneous effects of the CpG predictors   on the PCL scores.   In addition, we incorporate the numbers of  traumas and stressful events as homogeneous control variables.   The  DNHS has $126$ individuals with traumas whose average PCL scores in the first and second waves are completely observed. Since missing rates of average PCL scores from the third to fifth waves are higher than $50\%$ and our sample size is limited,    we impute the missing response values $y^*_{it}$ (for the $i$th individual at the $t$th wave)  from $N(\mu_i,0.35^2)$, where $\mu_i$ is the individual mean calculated  based on previous observed $y_{it}$'s, while  $0.35$ is determined  based on the   sample standard deviation of all complete responses.   We split the data into training and testing sets with three waves and two waves, respectively.

Given the limited number of individual-wise repeated measurements (three waves  for training) and the ultrahigh-dimensional covariates (1,648 CpG sites), we carry out  a screening process to identify potential covariates with significant  heterogeneous effects.  We fit a marginal homogenous model for each CpG predictor and filter out the CpG cites with p-values  greater than 0.4,  which are unlikely to have  significant effects for any reasonably large subgroup.   For  the remaining 376 covariates, we fit a marginal MDSP model to each of them and estimate the number of subgroups based on the gap statistic \citep{Tibshirani:2001}.  We are able to identify  three CpG sites (\textit{cg03256465,  cg03762702} and  \textit{cg06473843}) which  have significant heterogeneous effects.

For illustration, we compare the proposed MDSP model with the homogeneous regression model and the mixture-of-regression model \cite{McLachlan:2000}. Notice that all DNAm values at the  CpG sites  are measured only once, thus there is no variation on those  covariates within an  individual over longitudinal waves.  Therefore, any individual-wise models  such as  the individual-wise OLS model and the Lasso model as well as the random-effects model  are inapplicable.   We implement the mixture of regression model by the R package \textit{``mixtools''} (version 1.1.0) where the number of the mixture components is selected as  two by   bootstrap sequential testing \cite{McLachlan:2000}. 

To evaluate the model performance, we calculate the average prediction RMSE of the response  PCL scores   on the testing dataset. In addition, to examine whether subgrouping (the MDSP model and the mixture model) provides more informative data structure, we refit  a homogeneous model  within each identified subgroup,  and report the marginal p-values for CpG predictors, respectively. 

Table \ref{table:realdata} summarizes the RMSE values and the p-values of the  estimated CpG  coefficients. The MDSP model reduces the RMSE by 15\% and 32\% comparing to the mixture model and the homogeneous model, respectively.   
For  variable selection, the homogeneous model  does not provide  any significant results.   However,  the MDSP model successfully obtains  significant p-values corresponding to three  CpG sites  with  identified non-zero-effect subgroups, while the p-values in the zero-effect subgroups are clearly insignificant.  In contrast,  only one CpG site (\textit{cg0647384}) presents  significance in one subgroup of the mixture model (Component 1).   This  indicates that the MDSP model provides more informative subgrouping structure as it achieves individualized   variable selection and  subgrouping simultaneously.     Additionally,  we note  that the non-zero-effect subgroups identified by the MDSP model have reasonably large sizes,  consisting of  36.5\%, 34.2\% and 40.4\% of sample size  with respect to CpG sites   \textit{cg03256465,  cg03762702}  and  \textit{cg06473843}.

In Section B.2 of the Supplementary Materials,  we provide another illustration of the proposed method  analyzing the Harvard longitudinal AIDS clinical trial group   data to  investigate the heterogeneous treatment effects of Zidovudine on CD4 cell counts.

\section{Discussion}

In this paper,  we consider  an individualized regression model where both the number of individuals and the number of individual-wise measurements increase.   To select unique  features  for different individuals, we propose a  novel multi-directional separation penalty to implement individualized variable selection. In addition, by utilizing subpopulation structure, we induce  sub-homogeneous effects  and   borrow  cross-individual information to achieve a good balance of  parsimonious modeling  and heterogeneous  interpretation.

In contrast to  conventional penalized variable selection approaches,  the proposed method  provides multiple shrinking directions to overcome  the estimation bias  from  convex penalizations,  which prevent  strong signals being mistakenly pulled towards zero while pursuing model sparsity.   The alternative shrinking directions in addition to zero are automatically  selected  as potential subgroup effects through grouping of individuals with similar effects from predictors.   
Moreover, by incorporating within-individual serial correlation, the proposed method is able to gain  more efficiency than the model assuming independence.

In subgroup analysis,  to access  heterogeneous covariates'  effects, the existing  literature   \citep{Shuster:1983, Gail:1985,Yusuf:1991, Lagakos:2006, WangR:2007, Gunter:2011,Rendle:2012} proposes   adding   more interaction terms under a homogeneous model setting, which relies  on  pre-specified model assumptions such as   linear relationships  \citep{Gunter:2011,Rendle:2012}.  However, these assumptions are usually difficult to verify in applications.      The  covariates' heterogeneity could be more complex due to, for example,  unobserved factors rather than observed covariates.  By contrast, the proposed method  detects heterogeneous structures on individual covariates'  effects without relying on additional model assumptions on subgroup mechanisms.

To provide  individual-wise model inference, we lay out a double-divergence theoretical  framework which allows both sample size and individual-wise measurement size to diverge, and  also incorporates a divergent longitudinal correlation structure. The established large sample results indicate that  the proposed method achieves  a strong oracle property and thus  inherits the optimal convergence rate with true subpopulation information.  In addition, we also provide the optimal divergence rate of the dimension of individualized parameters as the  sample size increases.

In this paper, the individualized and the population-shared predictors are pre-specified in the model.   Therefore,  it is also essential to develop a method  to  test individualized variables from   population-shared  variables rather than depending on subgroup number selection.   
In addition,  we currently  assume a fixed number $p$ of individualized predictors, which can be   extended to a high-dimensional setting in which $p$ is also diverging.   This extension can  basically follow the standard results for a high-dimensional setting applying  on an individual-wise Lasso model, and then incorporating  grouping effects through a similar strategy  as in proving Theorem \ref{cthm:beta} in this paper.

%\begin{flushleft}\large
% \textbf{Acknowledgments}\\
%\end{flushleft}
%The authors would like to acknowledge support for this project from the National Science Foundation grants DMS-1613190, DMS-1821198, the National Institute of Health grant 7R01MD011728-03, and Dr. Monica Uddin to provide the PTSD data.

\begin{spacing}{1.2}

\end{spacing}

%%%%%%%%%%%%%%%%%%%%%Tables%%%%%%%%%%%%%%%%%%%%%%%%%%%%%

%%%%%%%%%%%%%%%%%%%%%%%%%%%%%%%%%%Simulation 1 Case a###############################################################
\vspace{0.5in}
\begin{table}[H]
\centering
\caption{\small The average RMSE of the proposed MDSP model compared with other approaches based on 100 simulations, with sample size $N=40,100$, cluster size (individual measurement size) $m=10, 20$  where Sub, Homo, FusedL, Lasso, AdapL, SCAD and MCP stand for individual-wise model, homogeneous model, the fused Lasso, the Lasso, the adaptive Lasso, the SCAD and the MCP  regularization models, respectively. } \label{table:s1armse}
\begin{tabular}{rrcccccccc}
\toprule
 {Sample }& {Cluster} &         \multicolumn{8}{c}{Methods} \\
{Size (N)}& {Size(m)} &  \textbf{MDSP} & Sub    & Homo & FusedL   &  Lasso & AdapL & SCAD & MCP   \\ \hline
                   &  &\multicolumn{7}{c}{$\gamma=1$} \\
\multirow{2}{*}{40}    &  $10$    & 0.267   & $0.349$     &$0.504$ & $0.323$   & $0.439$ &  $0.339$  & $0.344$ &0.350\\
                       &  $20$    & \textbf{0.120}   & $0.232$     &$0.502$ & $0.206$   & $0.298$ &  $0.207$  & $0.201$ &0.201\\
\multirow{2}{*}{100}   &  $10$    & 0.262   & $0.350$     &$0.501$ & $0.319$   & $0.394$ &  $0.334$  & $0.335$ &0.345\\
                       &  $20$    & \textbf{0.119}   & $0.233$     &$0.501$ & $0.210$   & $0.271$ &  $0.208$  & $0.205$&0.206\\\hline
                   &  &\multicolumn{7}{c}{$\gamma=2$} \\
\multirow{2}{*}{40}    &  $10$    & \textbf{0.122}   & $0.349$     &$1.004$ & $0.317$   & $0.408$ &  $0.309$  & $0.311$ &0.309\\
                       &  $20$    & \textbf{0.048}   & $0.232$     &$1.002$ & $0.204$   & $0.293$ &  $0.181$  & $0.168$ &0.167\\
\multirow{2}{*}{100}   &  $10$    & \textbf{0.113}   & $0.350$     &$1.001$ & $0.318$   & $0.387$ &  $0.305$  & $0.300$ &0.299\\
                       &  $20$    & \textbf{0.037}   & $0.233$     &$1.001$ & $0.210$   & $0.274$ &  $0.208$  & $0.206$ &0.206\\
\bottomrule
\end{tabular}
\end{table}

%%%%%%%%%%%%%%%%%%%%%%%%%%%%%%%%%%%%%%%Simulation  1 Case b %%%%
\begin{table}[H]
\centering
\caption{The average root mean square error (RMSE) of the proposed MDSP model with different working correlation structures based on 100 simulations, including AR-1 ($\bm{\beta}_{AR1}$), exchangeable ($\bm{\beta}_{Ex}$) and independent ($\bm{\beta}_{Ind}$) models. The true structures for the within-individual serial correlation are AR-1 or exchangeable, and correlation parameter $\rho=0.5$, sample size $N=20, 80$, cluster size (individual measurement size) $m=10, 20$. }
\label{table:s1brmse}
\begin{tabular}{ccrrrrrrr}
\toprule
  True              &  Cluster    &  \multicolumn{3}{c}{$N=20$} & & \multicolumn{3}{c}{$N=80$} \\
Correlation         &  size (m) &  $\bm{\beta}_{AR1}$ & $\bm{\beta}_{Ex}$ & $\bm{\beta}_{Ind}$ & &$\bm{\beta}_{AR1}$ & $\bm{\beta}_{Ex}$ & $\bm{\beta}_{Ind}$ \\\hline
  \multirow{2}{*}{Exch}             & 10          & 0.209 & \textbf{0.165} & 0.265 & &0.193 & \textbf{0.110} & 0.258 \\
                    & 20          & 0.072 & \textbf{0.053} & 0.078 & &0.067 & \textbf{0.051} & 0.076  \\
                    &             &       &  &  &  &  & & \\
  \multirow{2}{*}{AR-1}             & 10          & \textbf{0.182} & 0.230 & 0.258 & &\textbf{0.183} & 0.205 & 0.256 \\
                    & 20          & \textbf{0.091} & 0.121 & 0.132 & &\textbf{0.089} & 0.112 & 0.130  \\
\bottomrule
\end{tabular}
\end{table}

%%%%%%%%%%%%%%%%%%%%%%%%%%%%%%%%%%%%%%%%%%%%%%%%%%%%%%%%Simulation 2 %%%%%%%%%%%%%%%%%%%%%%%%%%%%%%%%%%%%%%%%%%%%%%%%%%%%%%%%%%%%%%%%%

\begin{table}[H]
\centering
\footnotesize
\caption{The average RMSE and CVSR of the proposed MDSP model compared to the individual-wise model (Sub), the fused Lasso (FusedL), the Lasso, the adaptive Lasso (Adapl), the SCAD and the MCP penalization models, with sample size $N=60$ and cluster size (individual measurement size) $m=10$. Scenario 1 contains a population homogeneous effect ($G_{k}=1$) and Scenario 2 contains an individualized predictor of three subgroups ($G_{k}=3$) with equal subgroup size.  In both cases the MDSP model assumes two subgroups, where the estimated sub-homogeneous effects are $\hat{\gamma}=2.01(0.06)$ and $\hat{\gamma}=-2.99(0.06)$ (with empirical standard errors in parenthesis), respectively.}
\label{table:s2misp}
\begin{tabular}{cclcccccc}
\toprule
Scenario & &  \textbf{MDSP}  & Sub  &FusedL   &  Lasso & AdapL & SCAD  & MCP  \\ \hline
{\footnotesize $G_{k}=1$}&RMSE            &0.115&0.346&0.319&0.414&0.373&0.346&0.345\\
{\footnotesize ($\beta_i=2$})&CVSR             &0.996&-&0.993&0.994&0.992&0.995&0.996\\
&&&&&&&&\\
{\footnotesize $G_{k}=3$ }&RMSE          &0.277&0.349&0.315&0.410&0.335&0.337&0.338\\
{\footnotesize ($\beta_i=-3,0,1$)}&CVSR     &0.901&-&0.748&0.877&0.902&0.816&0.817\\
\bottomrule
\end{tabular}
\end{table}

%%%%%%%%%%%%%%%%Real Data%%%%%%%%%%%%%%%%%%%%
\begin{table}[H]
\centering

\caption{The p-values of the estimated CpG coefficients in DNHS study from the homogeneous model,  the refitted model within subgroups identified by the MDSP model ($\mathcal{G}^{(0)}$ and $\mathcal{G}^{(\gamma)}$),  and by the mixture model (Comp 1 and Comp 2), and the prediction RMSE of PCL scores on testing set. }
\label{table:realdata}
\begin{tabular}{lccccc}
\toprule
                & \multicolumn{5}{c}{P-values of the coefficients}  \\ \cline{2-6}
\multirow{2}{*}{CpG sites}          & Homogeneous & \multicolumn{2}{c}{MDSP} & \multicolumn{2}{c}{MixReg} \\
                &             & $\mathcal{G}^{(0)}$    & $\mathcal{G}^{(\gamma)}$ (Proportion)   & Comp1    & Comp2\\ \hline
cg03256465      & 0.189       & 0.708    & \textbf {0.001} (36.5\%)   &0.783&  0.228    \\
cg03762702      & 0.396       & 0.468    & \textbf {0.029} (34.1\%) &0.189&   0.223   \\
cg06473843      & 0.376       & 0.156    & \textbf {0.001} (40.4\%)   &0.007& 0.082     \\\hline \hline
Prediction RMSE & 0.385       & \multicolumn{2}{c}{\textbf {0.292}}& \multicolumn{2}{c}{0.336}\\
\bottomrule
\end{tabular}
\end{table}

%%%%%%%%%%%%%%%%%%%%%figures %%%%%%%%%%%%%%%%%%%%%%%%%%%

\begin{figure}[H]
\begin{subfigure}{.35\textwidth}
\centering
\includegraphics[width=0.8\linewidth, height=3in]{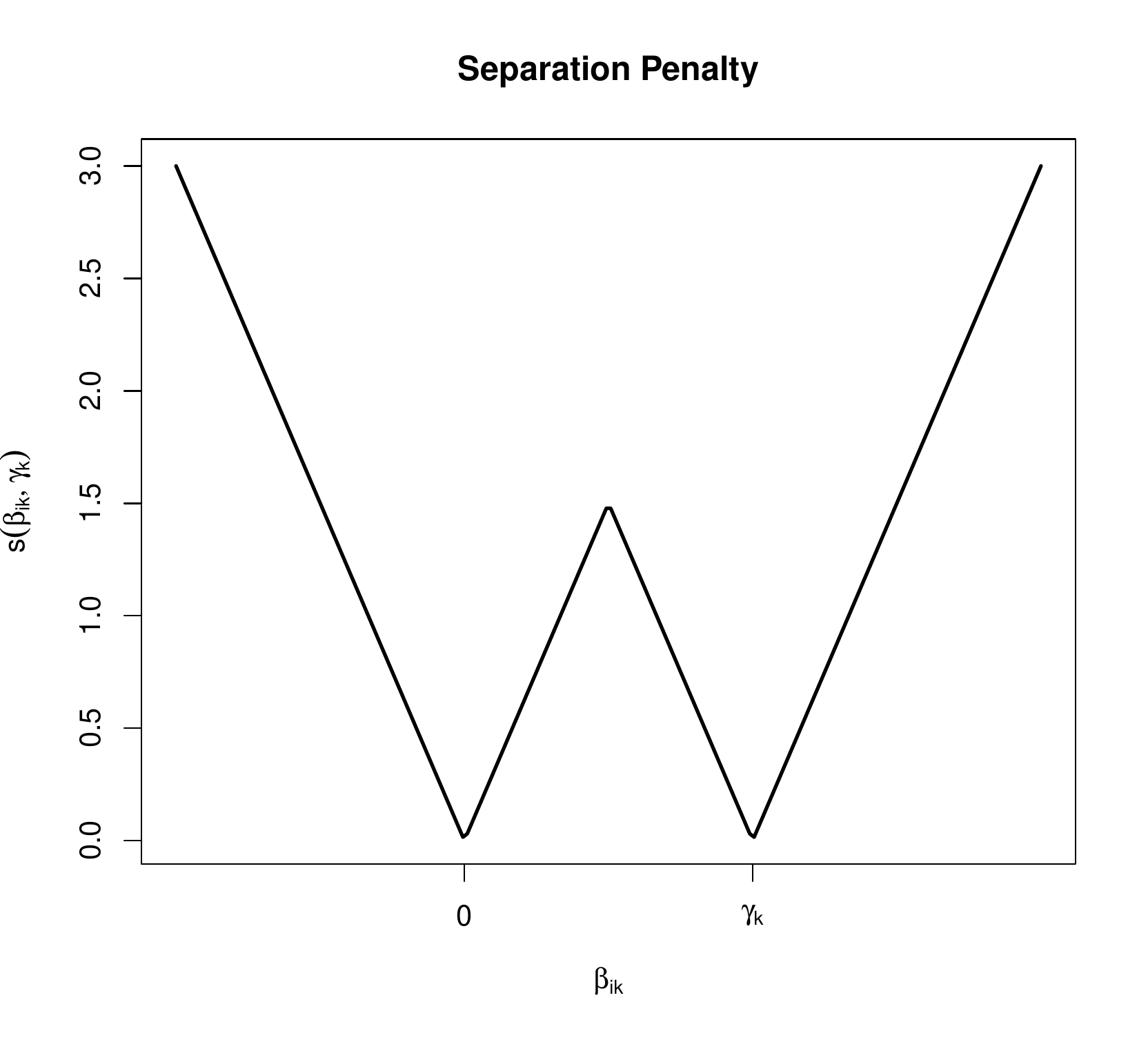}
\captionsetup{font=small}
\caption{\small The MDSP function $s(\cdot, \gamma_k)$.}
\label{fig:penalty}
\end{subfigure}
\begin{subfigure}{.65\textwidth}
\centering
\includegraphics[width=1\linewidth, height=3in ]{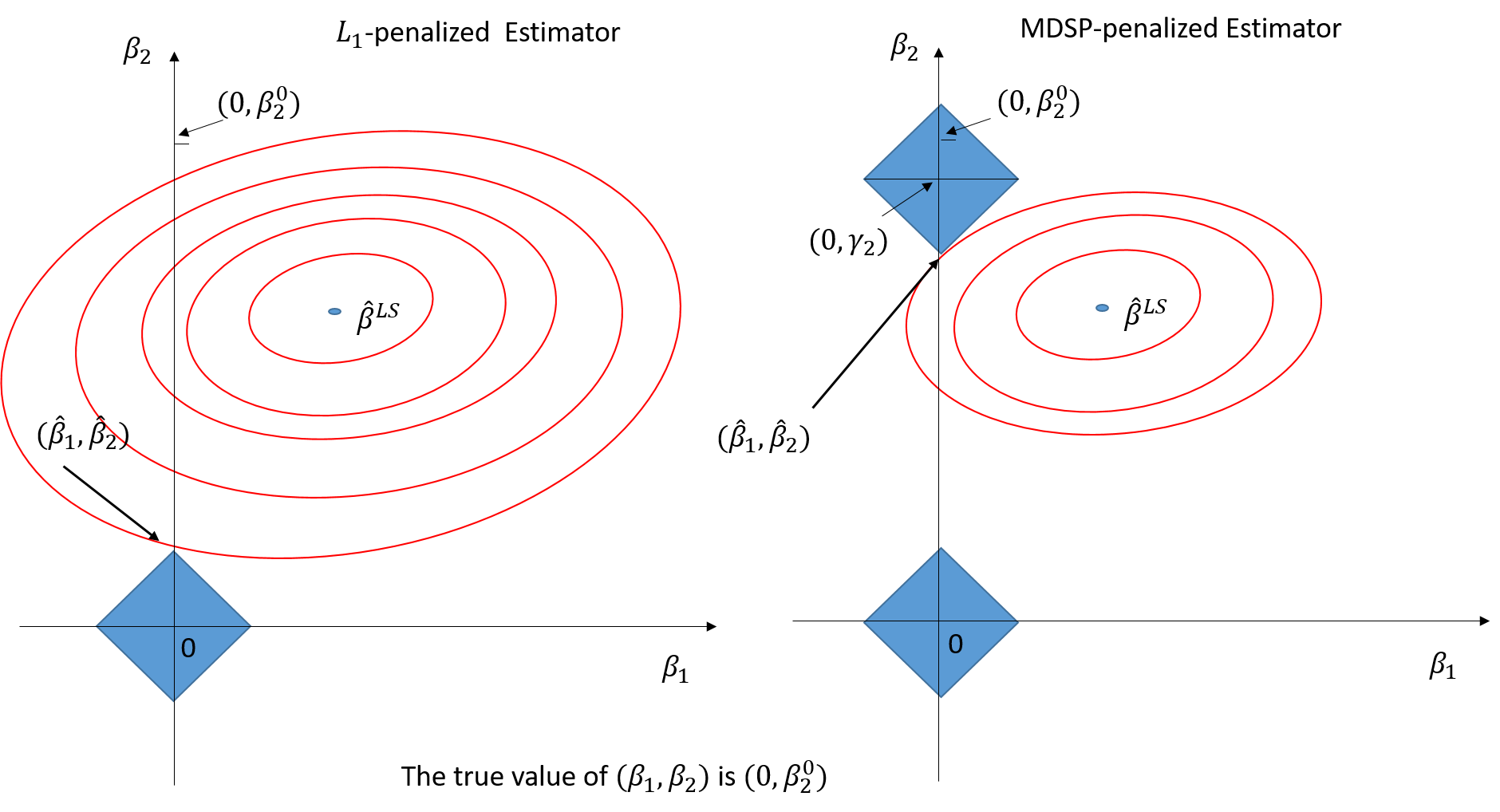}
\caption{\small  The $L_1$-penalized  estimator ($\sum_{j=1}^2 |\beta_j|$) and the MDSP-penalized  estimators ($\sum_{j=1}^2 \min(|\beta_j|, |\beta_j -\gamma_j|)$) for a subpopulation where the true value of $(\beta_1, \beta_2)$ is $(0,\beta_{2}^0)$, and $\hat{\bm{\beta}}^{LS}$ denotes the OLS estimator.}
\label{fig:mdsp}
\end{subfigure}
\caption{Illustration of the MDSP function and the MDSP-penalized estimators. }
\end{figure}

\newpage
\begin{figure}[H]
    \centering
    \includegraphics[width=1\linewidth, height=8cm]{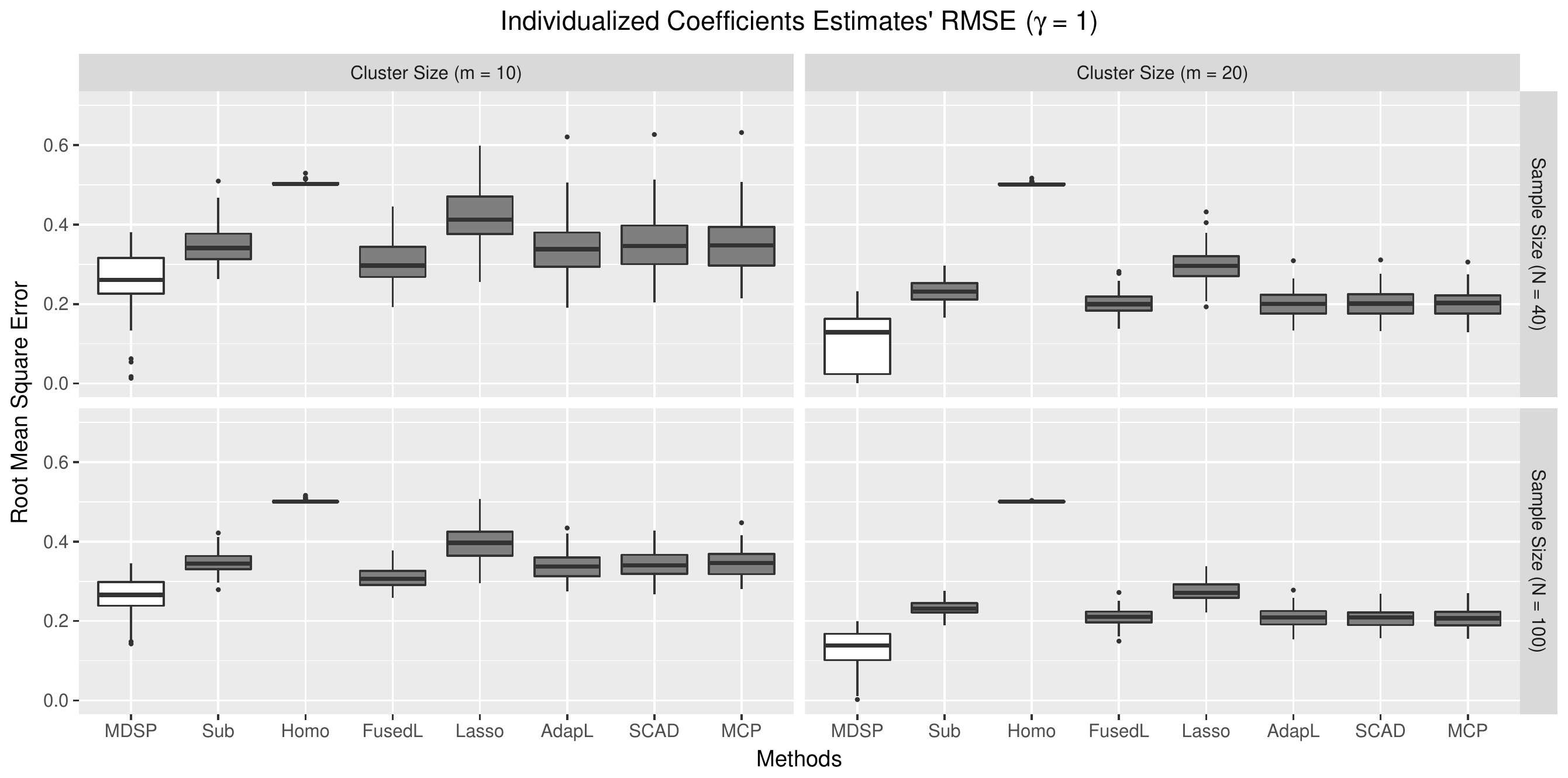}
    \caption{\small  The boxplot of RMSE of the proposed MDSP model compared with other approaches based on 100 simulations, with sample size $N=40,100$, individual measurement size (cluster size) $m=10, 20$, where homogeneous effect $\gamma=1$.}
        \label{fig:s1armse1}
\end{figure}
\begin{figure}[H]
    \centering
    \includegraphics[width=1\linewidth, height=8cm]{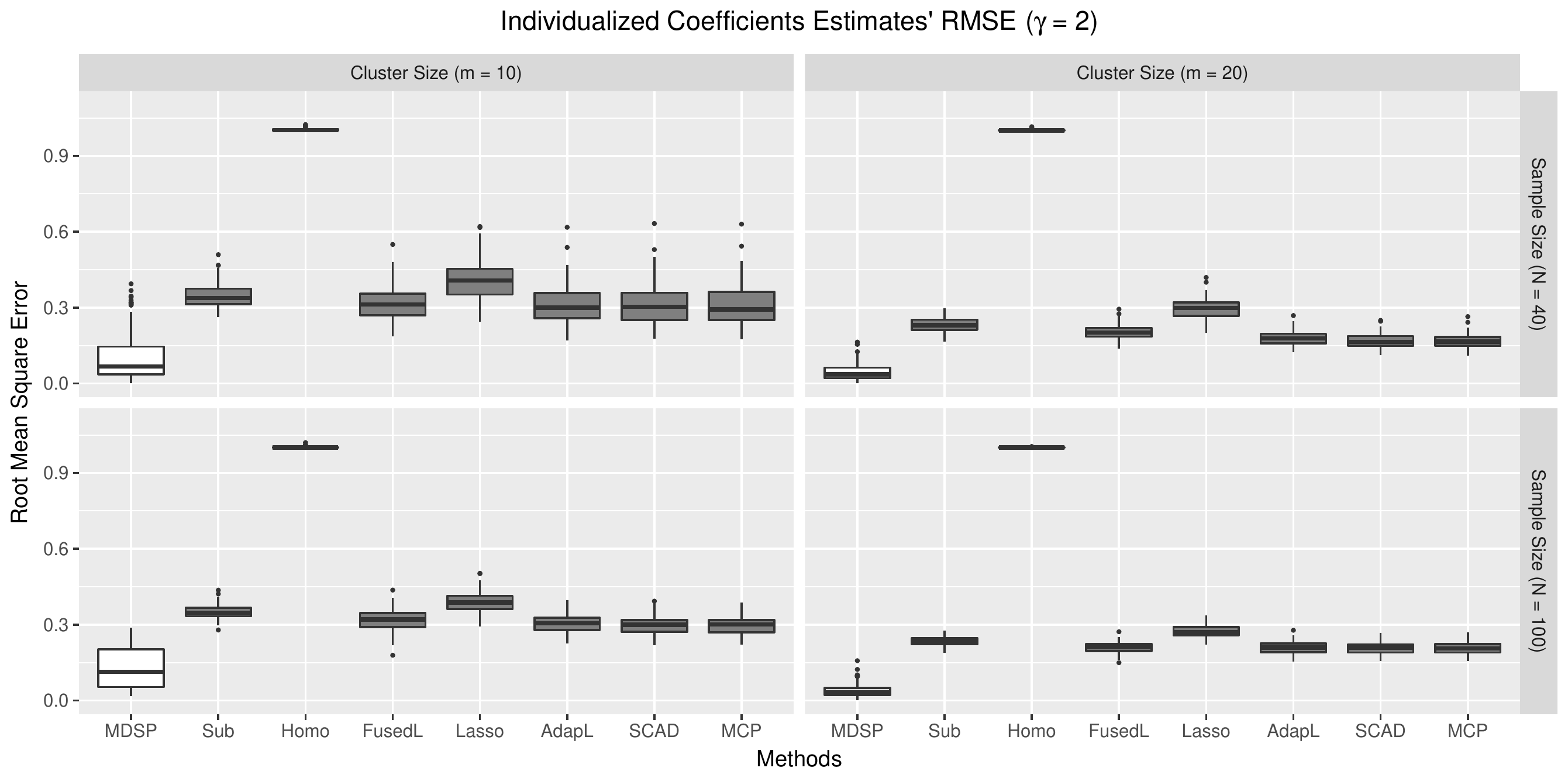}
        \caption{\small  The boxplot of RMSE of the proposed MDSP model compared with other approaches based on 100 simulations, with sample size $N=40,100$,  individual measurement size (cluster size) $m=10, 20$, where homogeneous effect $\gamma=2$.}
        \label{fig:s1armse2}
\end{figure}

\newpage

\begin{figure}[H]
    \centering
    \includegraphics[width=1\linewidth, height=8cm]{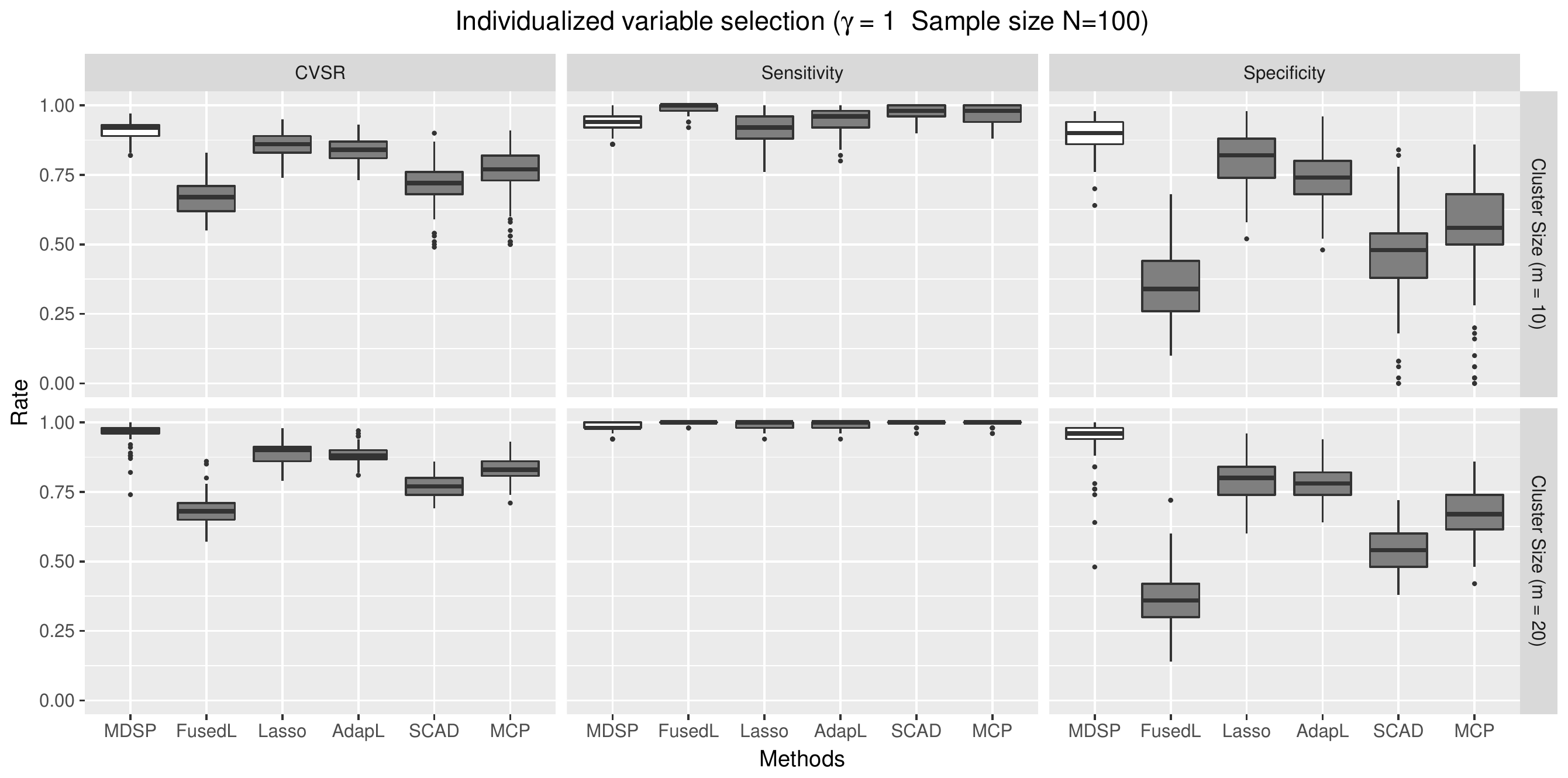}
    \caption{\small  The boxplots of CVSR, sensitivity and specificity for  all regularization approaches based on 100 simulations, with individual measurement size (cluster size) $m=10, 20$, where homogeneous effect $\gamma=1$ and sample size $N=100$.}
    \label{fig:s1avs2}
\end{figure}

\begin{figure}[H]
    \centering
    \includegraphics[width=1\linewidth, height=8cm]{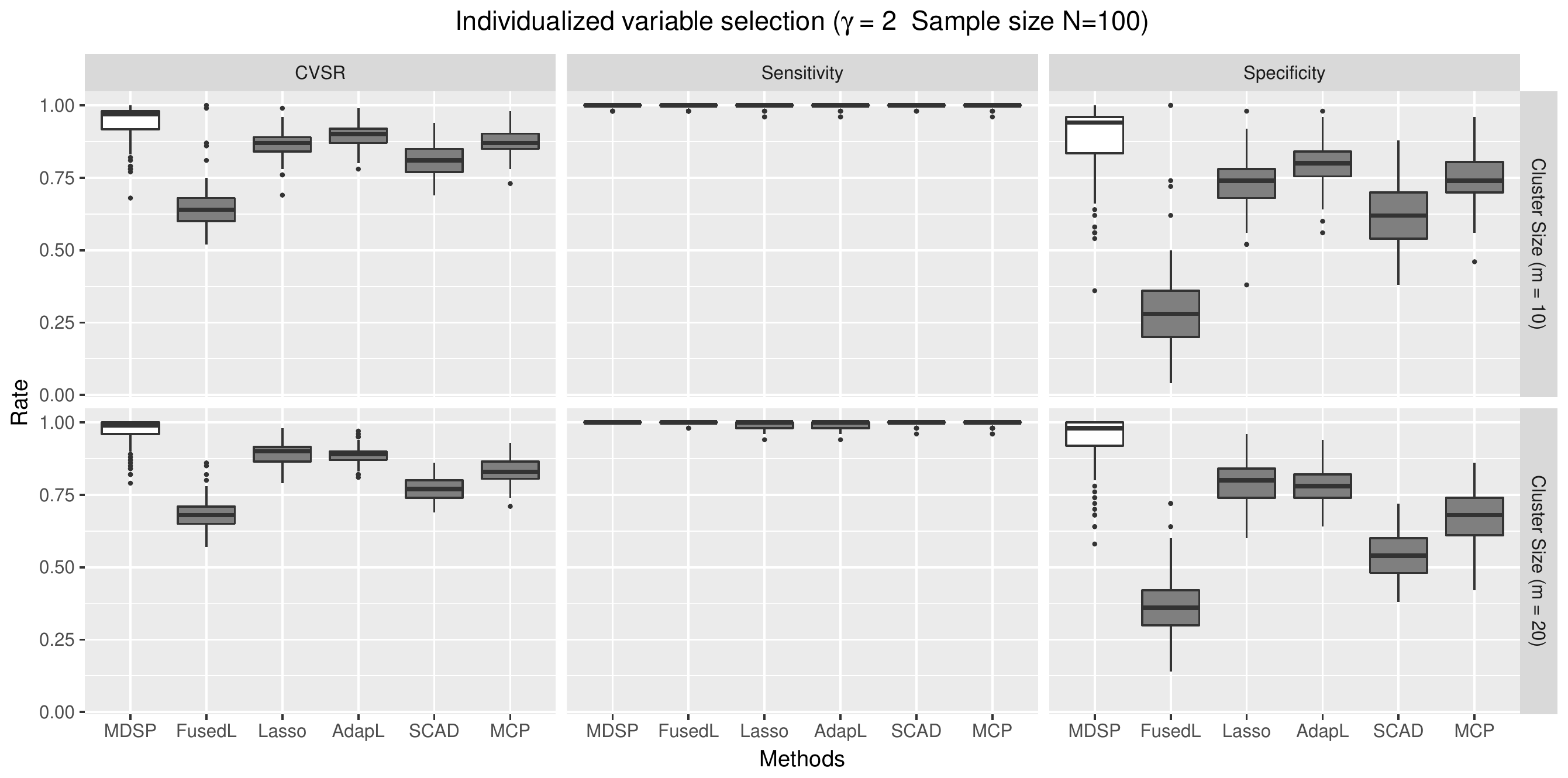}
    \caption{\small  The boxplots of CVSR, sensitivity and specificity for  all regularization approaches based on 100 simulations, with individual measurement size (cluster size) $m=10, 20$, where homogeneous effect $\gamma=2$ and sample size $N=100$.}
    \label{fig:s1avs4}
\end{figure}

\begin{figure}[H]
\includegraphics[width=1\linewidth,height=2.5in]{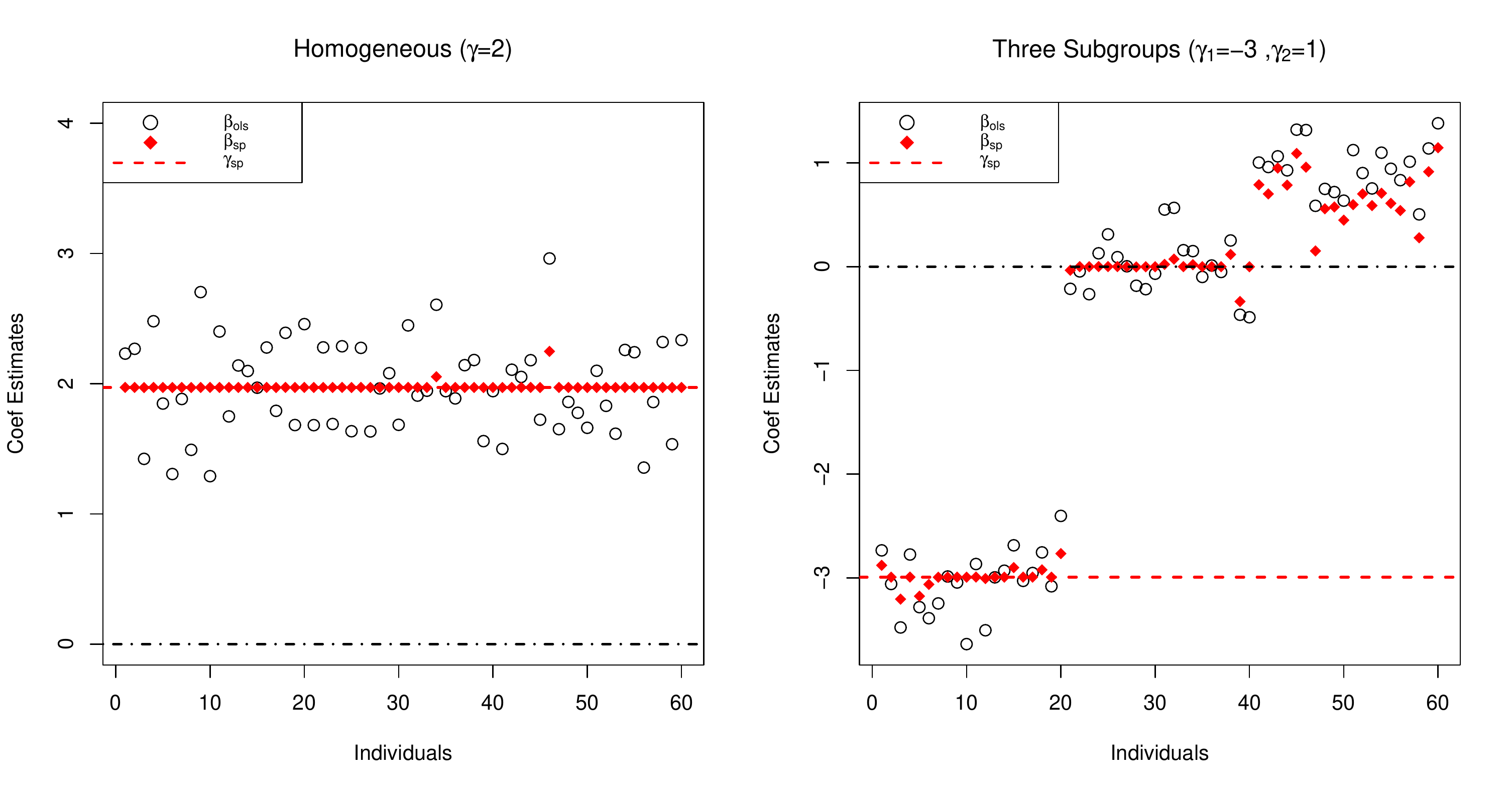}
\caption{\small The individual-wise least squares estimator and the proposed estimator  assuming two subgroups (including a zero group) for individualized parameters in  two scenarios:  a homogeneous group, and three subgroups,  where  the sample size $N=60$ and individual measurement size $m=10$. %where the proposed model both assumes  two subgroups (including zero group).
}
\label{fig:missp}
\end{figure}
\vspace{1in}
\begin{figure}[H]
    \includegraphics[width=1\linewidth]{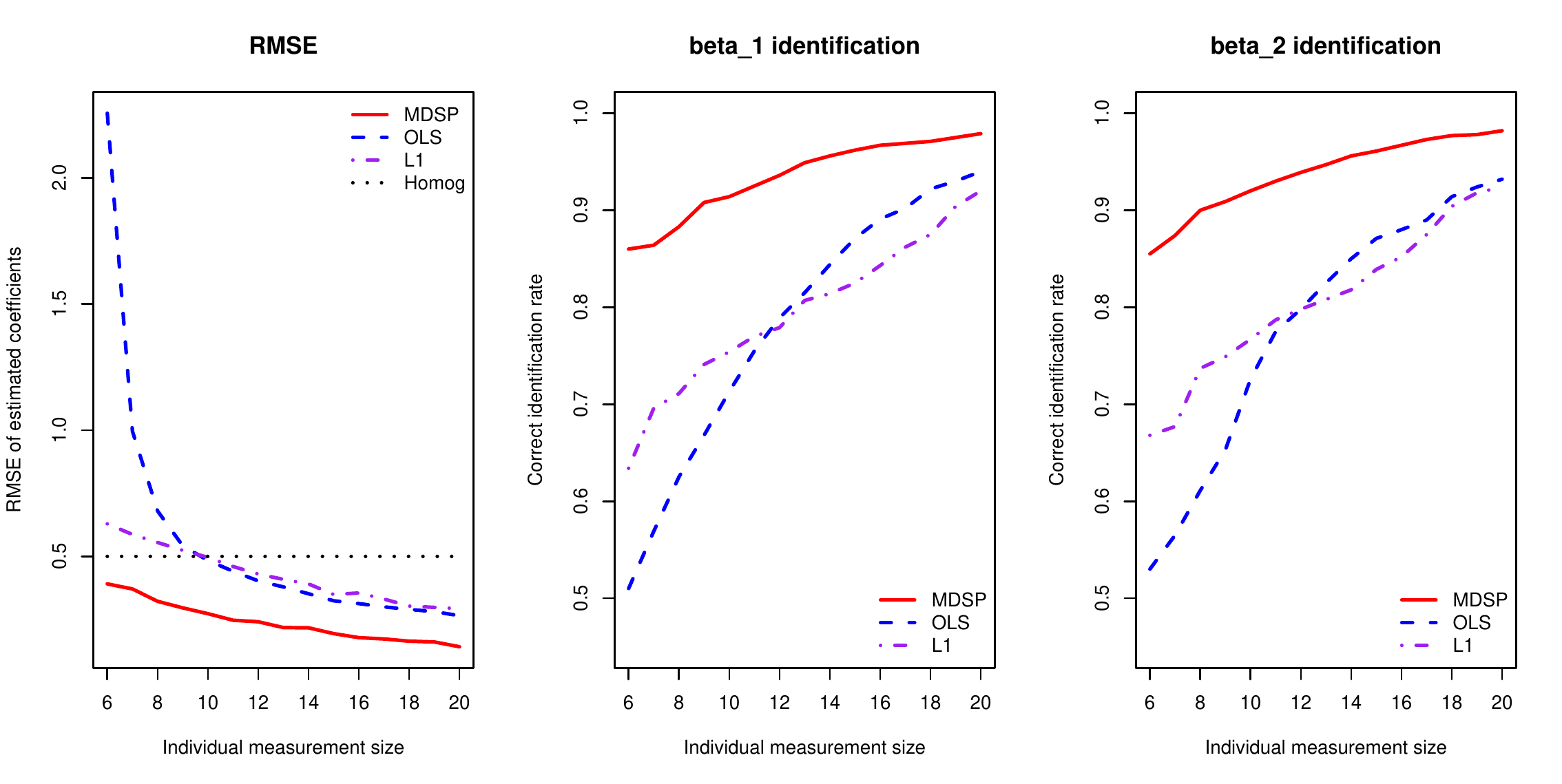}
    \caption{\small The left figure provides the average RMSE values of the coefficients estimations  ($(\hat{\beta}_1, \hat{\beta}_2)$ for the MDSP model, the individual-wise OLS model, the individual-wise Lasso (L1) model and the homogeneous model estimated on the training set.  The right two figures report the correct variable selection/elimination rates for $\beta_1$ and $\beta_2$, respectively.  All results are evaluated based on 5 replications of  $N^*=100$ semi-new individuals over different numbers of individual measurements ranging from 6 to 20. }
    \label{fig:newsub}
\end{figure}

\end{document}